\documentclass[12pt]{iopart}

\newcommand{\ket}[1]{|{#1}\rangle}
\newcommand{\bra}[1]{\langle{#1}|}
\newcommand{\eqref}[1]{(\ref{#1})}

\usepackage{iopams}

\usepackage{times}
\usepackage{graphicx}
\usepackage{subfigure}
\usepackage{hyperref} 
\usepackage{setstack}
\usepackage{cite}
\usepackage{color}

\begin{document}
\title{Ballistic propagation of a local impact in the one-dimensional $XY$ model}
\author{Atsuki Yoshinaga}
\ead{yoshi9d@iis.u-tokyo.ac.jp}
\address{Department of Physics, The University of Tokyo,
5-1-5 Kashiwanoha, Kashiwa, Chiba 277-8574, Japan}
\vspace{10pt}
\begin{abstract}
Light-cone-like propagation of information is a universal phenomenon of nonequilibrium dynamics of integrable spin systems.
In this paper, we investigate propagation of a local impact in the one-dimensional $XY$ model with the anisotropy $\gamma$ in a magnetic field $h$ by calculating the magnetization profile.
Applying a local and instantaneous unitary operation to the ground state, which we refer to as the local-impact protocol, we numerically observe various types of light-cone-like propagation in the parameter region $0\leq\gamma\leq1$ and $0\leq h \leq2$ of the model.
By combining numerical integration with an asymptotic analysis, we find the following:
(i) for $|h|\geq|1-\gamma^{2}|$ except for the case on the line $h=1$ with $0<\gamma<\sqrt{3}/2$, a wave front propagates with the maximum group velocity of quasiparticles, except for the case $\gamma=1$ and $0<h<1$, in which there is no clear wave front;
(ii) for $|h|<|1-\gamma^{2}|$ as well as on the line $h=1$ with $0<\gamma<\sqrt{3}/2$, a second wave front appears owing to multiple local extrema of the group velocity;
(iii) for $|h|=|1-\gamma^{2}|$, edges of the second wave front collapses at the origin, and as a result, the magnetization profile exhibits a ridge at the impacted site.
Furthermore, we find by an asymptotic analysis that the height of the wave front decays in a power law in time $t$ with various exponents depending on the model parameters:
the wave fronts exhibit a power-law decay $t^{-2/3}$ except for the line $h=1$, on which the decay can be given by either $\sim t^{-3/5}$ or $\sim t^{-1}$; the ridge at the impacted site for $|h|=|1-\gamma^{2}|$ shows the decay $t^{-1/2}$ as opposed to the decay $t^{-1}$ in other cases.
\end{abstract}

\section{Introduction}
Non-equilibrium dynamics of quantum many-body systems has been of great  theoretical and experimental interest.
Recent experimental and numerical advances in simulating and examining quantum dynamics have motivated a wide range of studies on dynamics of isolated quantum systems \cite{RevModColloquiumNoneq,2008coldAtom}.
The problems of thermalization and information propagation in isolated quantum systems are fundamental issues in this field.

Important questions include how and under what conditions a pure initial state approaches to thermal equilibrium through unitary time evolution.
Intensive studies in the last two decades have made remarkable progress in understanding the condition and mechanism of the thermalization in isolated quantum systems 
\cite{kinoshita2006quantum,rigol2007GGE,IntegrableQuench2016Review,schreiber2015observation,nandkishore2015many}.
A large number of theoretical and experimental studies, including the early investigation by von Neumann \cite{von2010proof}, have shown that local observables generally relax to their steady values, which in most cases are described by a thermal ensemble \cite{PRLeth1985,Deutsch,PRLHalTasaki1998,CanonicalTypicality2006,Popescu_2006,rigol2008thermalization,GoldsteinSheldonandLebowitz2010,trotzky2012probing,kaufman2016quantum}.
On the other hand, understanding of non-equilibrium dynamics towards the seemingly thermal state has not been well established yet since the way of equilibration varies considerably among the systems, and even the generic equilibration timescale has been unsolved \cite{timescale2017Pintos,RefWorks:doc:5e0465cae4b05da0baffb539,nickelsen2019classical}.

Among the phenomena of non-equilibrium dynamics in isolated quantum systems, ballistic spreading of a signal, namely the light-cone dynamics, is a widely observed one.
Such dynamics has been studied in various ways.
The celebrated Lieb-Robinson bound \cite{lieb1972finite} imposes an upper bound on the velocity of propagation of a local disturbance in systems with short-range interactions, and several important problems have been solved by its application \cite{hastings2004lieb,LRexpCluster2006,hastings2007area}.
While this rigorous result and seminal works offer an intuitive explanation for the light-cone behavior \cite{bravyi2006lieb,BrunoNachtergaeleOgata,kliesch2013liebrobinson}, the propagation dynamics exhibits a wide variety of phenomena depending on the situation.
For instance, the actual velocity of information propagation depends not only on the local Hamiltonian as the Lieb-Robinson velocity does, but also on the band structure of the total Hamiltonian and the initial state \cite{XXZ2014lightcone, XY2018quench}.
Indeed, we show below that a local impact propagates much slower than in the velocity given by the Lieb-Robinson bound.
In some systems \cite{LRveloSchuch2011,Kormos2014Referee2TWO,Bertini2017Referee2TWO,LRemergenceSciPostDubai2017}, it is even found that the information propagates at a finite speed when the Lieb-Robinson velocity diverges.

The most common setup of the Hamiltonian and the initial state to study information propagation is a protocol that we refer to as the global quench, in which one prepares the ground state of a given Hamiltonian and suddenly and permanently changes (namely ``quenches'') global system parameters, such as the interaction strength and a magnetic field \cite{IntegrableQuench2016Review,statistical1970XY,igloi2000long}.
Light-cone-like propagation of information has been observed under this protocol in a wide range of systems regardless of their integrability, mainly by calculating two-point correlation functions, entanglement entropy, and out-of-time ordered correlations, and the importance of information propagation in the relaxation process has been discussed  \cite{kaufman2016quantum,Huse2013,bohrdt2017scrambling,calabrese2005evolution,cheneau2012light,richerme2014non}.
The global-quench protocol is also used to explain the dynamics and the speed of propagation in integrable systems, in terms of a quasiparticle picture in which a pair of correlated particles are emitted from each point on the chain after a global quench and propagate with the maximum group velocity of the quasiparticles \cite{calabrese2005evolution,Calabrese_2006, Alba7947}.

Considering inhomogeneous initial states is another way of investigating propagation dynamics.
A protocol that has been often considered in the last decades is to connect the edge of two chains in different phases and producing an initial state with a domain wall \cite{Antal1999,Ogata2002firstNonEq,Ogata2002secondDiffusion,Hunyadi2004XX,divakaran2011non,eisler2013full,Bertini2016Referee2THREE,Viti2016JointInhomo,LocalNonIntegXXZBiella2016,GHDPRXYoshimura2016,Fagotti2016Referee2FOUR,Piroli2017Referee2THREE,perfetto2017ballistic,Bertini2017Referee2TWO,Collura2018Referee2THREE,Fagotti2020Referee2FIVE,GHDPRXBorsi2020}.
In this case, the energy and magnetization as well as the correlation functions exhibit propagation dynamics unlike in systems with homogeneous initial states.
In the $XX$ spin chain and the transverse-field Ising chain, a universal behavior characterized by the scaling $t^{-1/3}$ of time dependence has been revealed \cite{Hunyadi2004XX,Viti2016JointInhomo,perfetto2017ballistic}, as well as a staircase structure of magnetization profiles \cite{Hunyadi2004XX,eisler2013full}.
In fact, similar scaling has been also found in the asymptotic behavior around the light-cone edge of the correlation functions in the global-quench setting \cite{XY2018quench,PRA2012BoseGasPropVelo,Bertini2016edge}, since the Airy function, which is characterized by the time dependence $t^{-1/3}$, is used in both cases.
Another inhomogeneous initial conditions considered in this context is a local excitation, or a so-called droplet-like initial state, in which a few sites of a homogeneous system is perturbed \cite{karevski2002scaling,jurcevic2014quasiparticle,zauner2015time,caputa2017quantum}.
In the $XXZ$ model, a light-cone dynamics with multiple wave fronts has been found under local excitations \cite{XXZ2012boundstate,fukuhara2013microscopic,liu2014quench}.
While it can provide a further insight into understanding propagation properties of quantum systems, such a locally excited situation has been addressed in few studies so far, compared to the global quenches.

In the present paper, we investigate propagation dynamics of perturbation in an integrable quantum spin chain after locally and only instantaneously disturbing the system, which we refer to as the local-impact protocol.
We first prepare the ground state of a given system, and then apply a unitary operation $U_{\rm loc}$ which acts on the state only locally and instantaneously. 
The state after our protocol is therefore written as $\ket{\psi(t)}=e^{-iHt}U_{\rm loc}\ket{\psi}$, with the initial state $\ket{\psi}$ being an eigenstate of $H$.
We refer to the localized unitary operation $U_{\rm loc}$ as a local impact since it can be described by an instantaneous change of local parameters of the Hamiltonian, {\it i.e.}, $H(t)=H+V\delta_{n,0}\delta(t)$ with $U_{\rm loc}=\mathrm{e}^{-iV\delta_{n,0}}$, where we define the position of the disturbed site as the origin $n=0$.
Applying a spin-flip operation to one or a few consecutive sites of ground states, which have been considered in several studies \cite{jurcevic2014quasiparticle,zauner2015time,caputa2017quantum}, can be viewed as an example of the local-impact protocol, whereas considering a product state in which all spins are aligned up except for one or a few sites with spin down \cite{karevski2002scaling} is in general not, because the product state with all spins up is not necessarily an eigenstate of the Hamiltonian.

Important characteristics of the present protocol as opposed to other quench protocols include the following two points: (i) we can observe a light-cone-like propagation of quasiparticles in terms of local observables unlike in global quenches, which generally yield no transport in the dynamics; (ii) the translational invariance and the integrability of the Hamiltonian are conserved after the local impact in contrast to a local quench where the Hamiltonian is locally changed permanently \cite{Bertini2017Referee2TWO,LocalNonIntegXXZBiella2016,Fagotti2016Referee2FOUR,torres2014local,LocalImpurityLuca2014,TorresHerreraPhysScr2015,Fagotti2015controlArXiveReferee2FOUR,Bertini2016Referee2FOUR,Bastianello2018Referee2FOUR,LocalImpurityBastianello2019}.

We specifically consider the spin-$1/2$ anisotropic $XY$ chain \cite{statistical1970XY,lieb1961two,statistical1970XYtwo,statistical1970XYthree,statistical1970XYfour,perk2009new} in a magnetic field, and calculate the dynamics of the magnetization .
We find that the model exhibits rich propagation dynamics of the wave front, such as the existence of a second wave front and power-law decay with several exponents depending on the model parameters.
For the asymptotic behavior of the wave fronts, the Airy function has been widely used in the previous studies for integrable systems \cite{XY2018quench,eisler2013full,Hunyadi2004XX, Bertini2016edge,Viti2016JointInhomo,perfetto2017ballistic,PRA2012BoseGasPropVelo}.
We perform an asymptotic analysis by generalizing the Airy scaling techniques, and demonstrate that it successfully captures the long-time behavior of the wave fronts in most cases.
We also show that this technique fails when the model reduces to the Ising model, or when the system is on the Ising transition line.

This paper is organized as follows.
In Sec.~\ref{sec:set}, after introducing the model, we derive an integral form of the magnetization change under the local-impact protocol and perform an asymptotic analysis to find the velocity of the propagation.
In Sec.~\ref{sec:res}, we present the phase diagram according to the inflection points of the dispersion relation, or the local extrema of the group velocity of quasiparticles and investigate the propagation dynamics by numerical integration.
In Sec.~\ref{sec:AiryNew}, we perform a more precise analysis on the asymptotic behavior of the magnetization change, and discuss the origin of an anomalous behavior which is observed in Sec.~\ref{sec:res}.
We conclude the paper in Sec.~\ref{sec:disc}, summarizing our findings and proposing future research.
We also provide appendices to show details of calculation.

\section{Analytic calculation of the time evolution of the magnetization change}\label{sec:set}
We consider the one-dimensional spin-$1/2$ antiferromagnetic $XY$ model described by the Hamiltonian
\begin{eqnarray}
H=\sum_{n=0}^{L-1} \left[(1+\gamma)S_n^xS_{n+1}^x +(1-\gamma)S_n^yS_{n+1}^y+ h S_n^z\right] ,\label{eq:H}
\end{eqnarray}
where $\{S^{x}_{n},S^{y}_{n},S^{z}_{n}\}$ are the spin-1/2 operators,
$L$ denotes the system size, $\gamma$ denotes the $XY$ anisotropy, and $h$ denotes the magnitude of a magnetic field.
We require the periodic boundary conditions $\vec{S}_{L}=\vec{S}_{0}$ and take the system size $L$ to be an even number in the diagonalization below.
In this study, we particularly investigate the dynamics in the parameter region of $0\leq h \leq2$ and $0\leq\gamma\leq1$.

We here use the ground state $\ket{\rm GS}$ for the initial state of our local-impact protocol, and specifically give the local impact $U_{\rm loc} =  \mathrm{e}^{i\theta S_{0}^{z}}$, namely a rotation over the angle $\theta$ of $\vec S_{0}$ around the $z$-axis.
Then we analyze the spatial propagation of the effect of the local impact on the state by calculating the dynamics of the magnetization in the $z$ direction at each site $n$, according to the original Hamiltonian \eqref{eq:H}.
We focus on the change of the local magnetization
\begin{eqnarray}
\Delta(S^{z}; n, t)  \equiv  \bra{\rm GS} U_{\rm loc}^{\dagger} S_{n}^{z}(t) U_{\rm loc}  \ket{\rm GS} - \bra{\rm GS} S_{n}^{z} \ket{\rm GS},\label{eq:Delta2}
\end{eqnarray}
where $S_{n}^{z}(t)$ denotes $\mathrm{e}^{iHt}S_{n}^{z}\mathrm{e}^{-iHt}$.
(We set $\hbar=1$ here and hereafter.)
The Jordan-Wigner transformation, which we introduce later, makes Eq.~\eqref{eq:Delta2} equal to the change of the fermion density at site $n$.
In the calculation of the propagation dynamics, we take the thermodynamic limit $L\rightarrow\infty$.

In this section we first give a brief summary of the diagonalization of the $XY$ model in one dimension under the periodic boundary condition.
After that we derive an integral expression of the magnetization change $\Delta(S^{z}; n, t)$ and perform an asymptotic analysis in order to discuss the velocity of the propagating wave fronts.
All the results on the propagation dynamics in this study also hold for the ferromagnetic $XY$ model.

\subsection{Diagonalization of the $XY$ model in one dimension}\label{subsec:digonal}
For the diagonalization of the Hamiltonian \eqref{eq:H}, we rewrite it in terms of spinless fermions by using the Jordan-Wigner transformation \cite{JW1928}, which is defined by
\begin{eqnarray}
S_{n}^{z}= c_{n}^{\dagger} c_{n} -\frac12, 
\,\,\,S_{n}^{+}=c_n^{\dagger}\mathrm{e}^{i\pi\sum_{m=0}^{n-1}c_m^{\dagger}c_m},
\,\,\,S_{n}^{-}=\mathrm{e}^{-i\pi\sum_{m=0}^{n-1}c_m^{\dagger}c_m}c_n,\label{eq:JWdef}
\end{eqnarray}
where the operators $c_{n} , c_{n}^{\dagger}$ obey the fermionic anti-commutation relations $ \{ c_n, c_m \}=\{ c_n^{\dagger},c_m^{\dagger} \}=0$, $\{ c^{\dagger}_n ,c_m \}=\delta_{nm}$.
We thereby obtain
\begin{eqnarray}
H = \sum_{n=0}^{L-1}&\left[ \frac{1}2\left( c_n^{\dagger} c_{n+1} + \gamma c_n^{\dagger} c_{n+1}^{\dagger} + h.c. \right)+ h \left(c_{n}^{\dagger} c_{n} -\frac12 \right)\right],\label{eq:HamInC}
\end{eqnarray}
where the boundary condition is given by
\begin{eqnarray}
c_{L}= \mathrm{e}^{i\pi N}c_{0}\,\,,\,\,\,\, c_{L}^{\dagger}= c_{0}^{\dagger}\mathrm{e}^{-i\pi N}\label{eq:BCc}
\end{eqnarray}
with $N=\sum_{m=0}^{L-1}c_m^{\dagger}c_m$.
The operator $\mathrm{e}^{i\pi N}$ has the eigenvalues $\pm1$ and commutes with the Hamiltonian \eqref{eq:HamInC}.

We can therefore block-diagonalize the Hamiltonian as
\begin{eqnarray}
 H &= P_{+} H^{+}P_{+} \oplus P_{-} H^{-} P_{-}
 \label{eq:HamTotal}
\end{eqnarray}
with
\begin{eqnarray}
 H^{\pm} \equiv& \frac{1}2\sum_{n=0}^{L-2}\left( c_n^{\dagger} c_{n+1} + \gamma c_n^{\dagger} c_{n+1}^{\dagger} + h.c. \right) + h\sum_{n=0}^{L-1} \left(c_{n}^{\dagger} c_{n} -\frac12 \right) \nonumber\\
 &\mp \frac{1}2 \left[  (c_{L-1}^{\dagger}c_{0}+\gamma c_{L-1}^{\dagger}c_{0}^{\dagger} ) +h.c. \right],
\label{eq:HamInJW}
\end{eqnarray}
where
\begin{eqnarray}
P_{\pm} \equiv \frac12(1\pm\mathrm{e}^{i\pi N})\,\,\,\,{\rm with}\,\,\,P_{+}+P_{-}=1
\label{eq:projection}
\end{eqnarray}
are the projection operators onto the respective blocks, which commute with $H$ and $H^{\pm}$ as well as $S_{n}^{z}$ for all $n$, and $h.c.$ denotes hermitian conjugate.
The blocks given by $P_{\pm}$ are sometimes referred to as the Neveu-Schwarz sector and the Ramond sector, respectively \cite{calabrese2012quantum}.

Using the Fourier transformation and the Bogoliubov transformation, we can diagonalize each of the Hamiltonians $H^{\pm}$ as
\begin{eqnarray}
H^{\pm}={\sum_{p}}^{\pm} \varepsilon(p) \left(\eta_{p}^{\dagger}\eta_{p} - \frac12\right),
\label{eq: HamiltonianPlusINeta}
\end{eqnarray}
where the fermion $\eta_{p}$, namely a quasiparticle defined by
\begin{eqnarray}
\eta_{p}  =\sqrt{\frac{1}{L}}\sum_{n=0}^{L-1} \mathrm{e}^{inp}\left(s_{p} c_{n} - t_{p}c_{n}^{\dagger}\right)\label{eq:etapsptp}
\end{eqnarray}
with
\begin{eqnarray}
s_{p} &\equiv \sqrt{\frac{\varepsilon(p)+\cos{p}+h}{2\varepsilon(p)}},\label{eq:sp}\\
t_{p} &\equiv i\frac{\gamma \sin{p}}{|\gamma \sin{p}|} \sqrt{\frac{\varepsilon(p)-(\cos{p}+h)}{2\varepsilon(p)}},\label{eq:sptp}
\end{eqnarray}
satisfies the standard anti-commutation relations $\{\eta_{p},\,\eta_{q}\}=\{\eta_{p}^{\dagger},\,\eta_{q}^{\dagger}\}=0\,\,, \,\, \{\eta_{p},\,\eta_{q}^{\dagger}\}=\delta_{p,q}$.
For the summation $\sum_{p}^{\pm}$ over momentum $p=2\pi j /L$, we take $j=-(L-1)/2, ... ,-1/2, 1/2, ... ,(L-1)/2$ for the Neveu-Schwarz sector $H^{+}$ and  $j=-L/2-1, ... ,-1,0,1, ... ,L/2$ for the Ramond sector $H^{-}$ both with even $L$ so that the anti-periodic or periodic boundary condition \eqref{eq:BCc} may be satisfied.

The dispersion relation $\varepsilon(p)$ of the quasiparticles in Eq.~\eqref{eq: HamiltonianPlusINeta} is given by
\begin{eqnarray}
\varepsilon(p) &\equiv \sqrt{( \cos{p} +h)^{2} + (\gamma \sin{p} )^{2}}\label{eq:varEpsilon}
\end{eqnarray}
for the anisotropic case $\gamma\neq0$.
For the isotropic case $\gamma=0$, it reduces to
 \begin{eqnarray}
 \varepsilon(p) =   \cos{p} +h,
 \label{eq:DispXX}
 \end{eqnarray}
and hence we have $s_{p}=1$ and $t_{p}=0$.
\begin{figure}
\begin{indented}\item[]
\subfigure[]{
\centering
\includegraphics[width=0.4\textwidth]{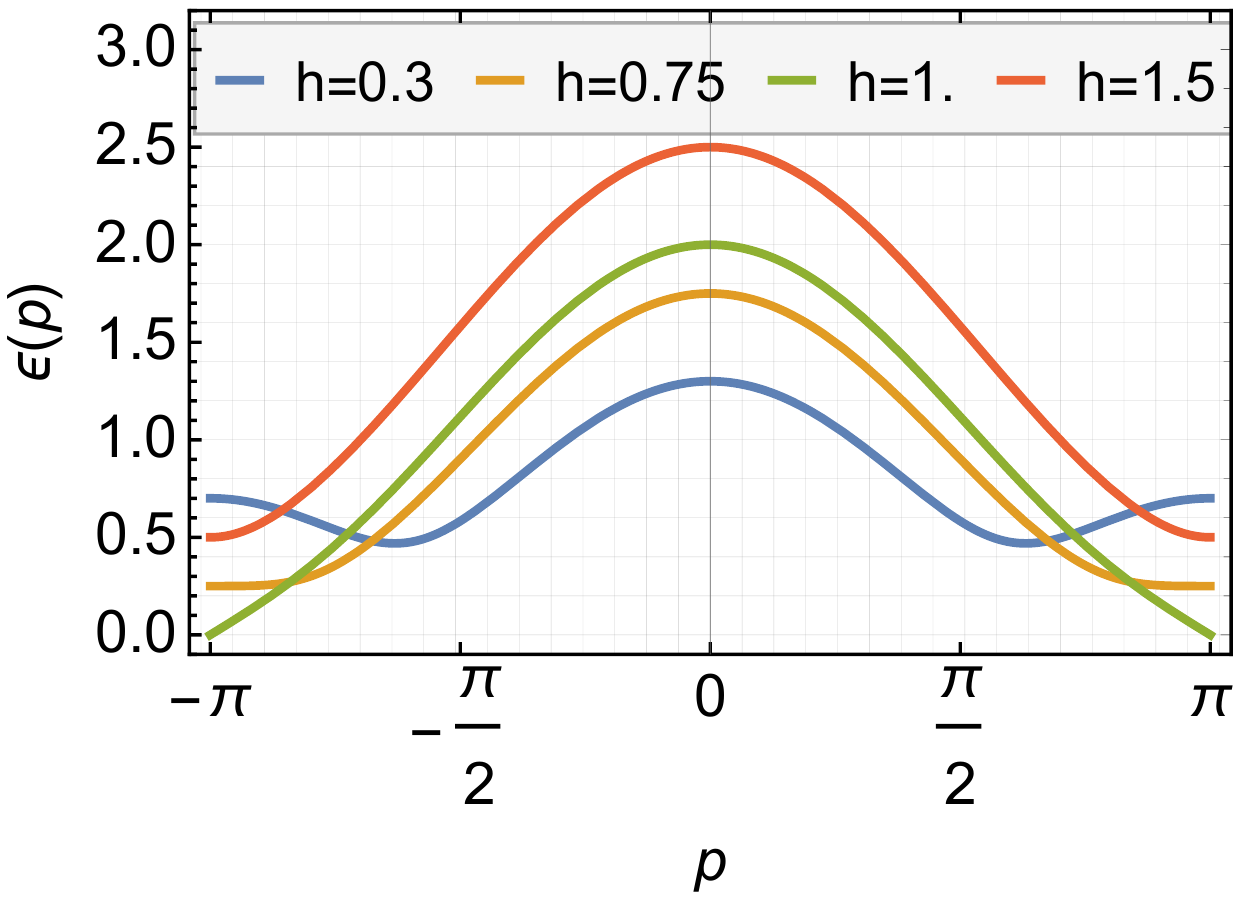}
\label{fig:DispersionG05}
}
\subfigure[]{
\centering
\mbox{\raisebox{3.5mm}{\includegraphics[width=0.415\textwidth]{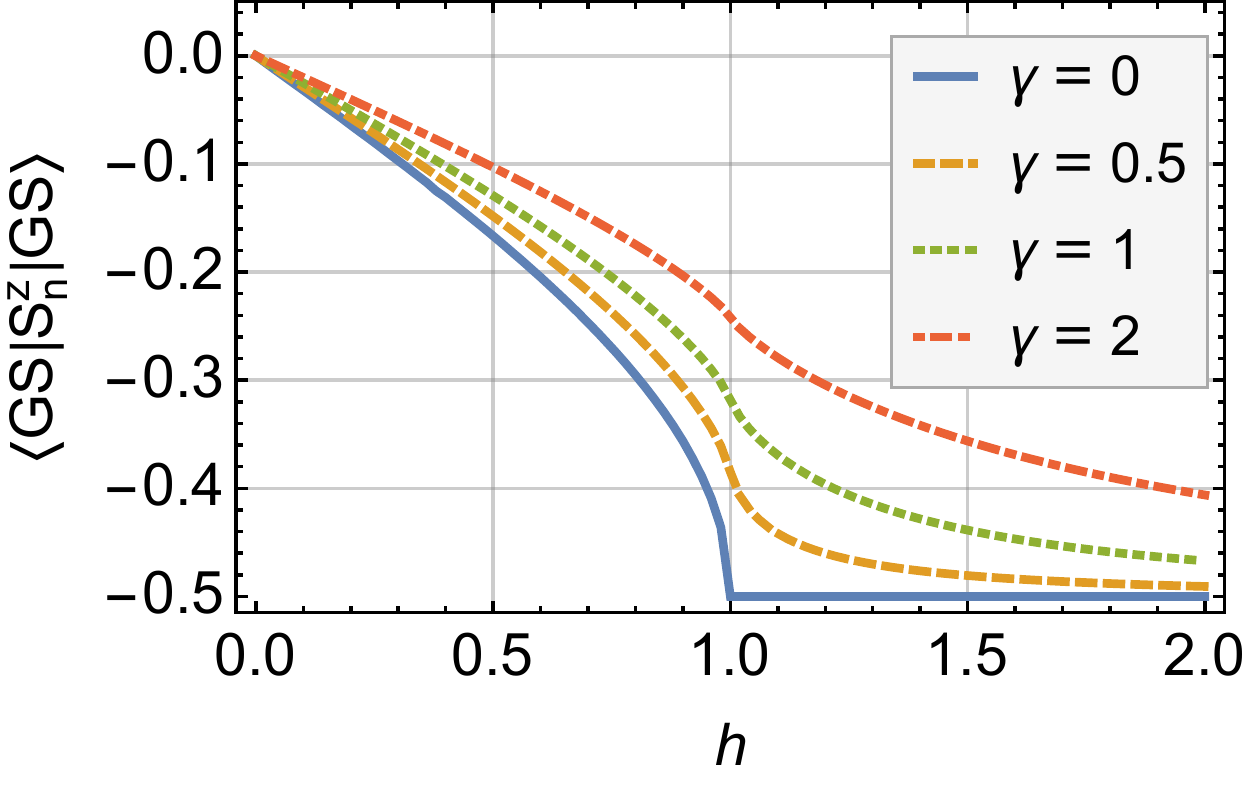}}}
\label{fig:SzGsVaryh}
}
\caption{The dispersion relation \eqref{eq:varEpsilon} and the ground-state magnetization along the $z$-axis of the $XY$ model.
(a) The dispersion relation in the first Brillouin zone for four values of the magnetic field $h$ with the anisotropy $\gamma=0.5$.
(b) The ground-state magnetization $\bra{\rm GS}S^{z}_{0}\ket{\rm GS}$ along the $z$-axis for four values of $\gamma$.
}
\label{fig:dispersionSzGsVaryh1}
\end{indented}
\end{figure}
The dispersion relation \eqref{eq:varEpsilon} can have a multimodal shape as we show in Fig.~\ref{fig:DispersionG05}.
The group velocity of the quasiparticles is given by
\begin{eqnarray}
v_{\rm g}(p)\equiv \frac{d}{dp}\varepsilon(p) = -\frac{\sin{p} \left(\left(1-\gamma^2\right) \cos{p}+h\right)}{\sqrt{(\cos{p}+h)^2+\gamma^2 \sin ^2{p}}}
\end{eqnarray}
for the anisotropic case $\gamma\neq0$, whereas $v_{\rm g}(p)=-\sin{p}$ for the isotropic case $\gamma=0$.

We here use the ground state of the Hamiltonian \eqref{eq:H} for the initial state of our local-impact protocol.
The ground state of Eq.~\eqref{eq:H} is given by either or both of the ground states $\ket{\rm GS}_{\pm}$ of the Hamiltonians \eqref{eq: HamiltonianPlusINeta}, where the sign of the subscript of the ground states corresponds to that of the superscript of the Hamiltonians.
In fact, the choice of the ground state of Eq.~\eqref{eq:H} depends on $L$, $\gamma$, and $h$ as discussed in Ref.~\cite{okuyama2015anomalous}.
Nevertheless, whether we choose $\ket{\rm GS}_{+}$, $\ket{\rm GS}_{-}$, or a superposition of them as the ground state of $H$, is irrelevant in the calculation of $\Delta(S^{z}; n, t)$ for $L\gg1$, which we show in \ref{sec:AppA}.

For brevity, we here describe the derivation of $\ket{\rm GS}_{+}$ only.
For the anisotropic case, the ground state of $H^{+}$ is given by the vacuum of $\eta_{p}$ since $\varepsilon(p)>0$ for a finite even $L$: 
\begin{eqnarray}
\eta_{p}\ket{\rm GS}_{+}=0 \,\,\,\,\,\,{\rm for\,\, any\,\,} p.\label{eq:gsXY}
\end{eqnarray}
For the isotropic case, the ground state is the state in which only the levels with negative energies are filled with fermions $\eta_{p}$:
\begin{eqnarray}
\eta_{p}\ket{\rm GS}_{+}=0 \,\,\,\,\,\,{\rm for\,\,} p \,\,{\rm with }\,\, \varepsilon(p)=\cos{p}+h>0,\label{eq:gsXX}
\end{eqnarray}
where we assumed for simplicity that no momentum $p$ satisfies $\varepsilon(p)=\cos{p}+h=0$.
If there is a value of $p$ with $\varepsilon(p)=0$ and the ground state has degeneracy owing to this zero-energy excitation, it would only make difference of $O\left(1/L\right)$ in the magnetization change \eqref{eq:Delta2}.
We use Eqs.~\eqref{eq:gsXY} and \eqref{eq:gsXX} in deriving Eqs.~\eqref{eq:Fdefint}--\eqref{eq:Qdefintxx} from Eqs.~\eqref{eq:defFGbraket} and \eqref{eq:defQWbraket} in Sec.~\ref{subsec:SettingLast}.

Finally we discuss the role of the local impact regarding the quasiparticle excitation on the ground state.
The local impact that we use here is expressed as follows in terms of $\eta_{p}$:
\begin{eqnarray}
U_{\rm loc}&=\,\mathrm{e}^{i\theta S_{0}^{z}}=\cos{\frac{\theta}2}+i\sin{\frac{\theta}2}\left(2c_{0}^{\dagger}c_{0}-1\right)\nonumber\\
&=\cos{\frac{\theta}2}-i\sin{\frac{\theta}2} + 2\sin{\frac{\theta}2} \frac{i}L\sum_{p,q}(s_{p}\eta_{p}^{\dagger} - t_{p}\eta_{-p})(s_{q}\eta_{q}+ t_{q}\eta_{-q}^{\dagger}).\label{eq:Ulocex}
\end{eqnarray}
The third term on the right-hand side of this expression represents the creation and annihilation of the quasiparticles.
When this term is applied to the ground state \eqref{eq:gsXY} for the anisotropic case $\gamma\neq0$, it excites all possible pairs of quasiparticles with momentum $-p$ and $q$ since the ground state is the vacuum.
For the isotropic case, this term is reduced to $ \left[ 2\sin{\theta/2} \right](i/L)\sum_{p,q} \eta_{p}^{\dagger}\eta_{q}$.
In this case, the local impact excites all possible pairs of quasiparticle excitation and hole on the ground state \eqref{eq:gsXX} since it is occupied by the quasiparticles below the Fermi level.

In both cases, the local impact excites quasiparticles with broad range of energies.
As a consequence, information of the local impact is ballistically transferred by quasiparticles of all possible momenta, and the fastest quasiparticles form propagating wave fronts of a light cone, regardless of the detail of the local impact.
Even if the impact is weak in the sense that $U_{\rm loc}\sim I$, {\it i.e.}, $\theta\sim0$, the above picture holds, and the excitation is not limited to low-energy quasiparticles.

\subsection{Time evolution of the magnetization change}\label{subsec:SettingLast}
Now we present an analytical expression of the magnetization change \eqref{eq:Delta2}:
\begin{eqnarray}
\fl
\Delta(S^{z}; n, t) &=\langle{U_{\rm loc}^{\dagger} S_{n}^{z} (t)U_{\rm loc} }\rangle_{} - \langle{S_{n}^{z}(0)  }\rangle_{}= 4 \sin{\frac{\theta}2} \Im \left[ \mathrm{e}^{-i\frac{\theta}2} \left(  K_{1} +\langle{ c_{0}^{\dagger} c_{0}}\rangle_{}K_{2} \right)\right],
\label{eq:DeltaSzForm}
\end{eqnarray}
where the angular bracket $\langle\cdots \rangle$ denote the expectation value with respect to the ground state of our choice and
\begin{eqnarray}
K_{1}\equiv  FQ^{*}-GW^{*} \,,\,\,\,\,\,\,K_{2}\equiv |G|^{2} - |F|^{2}
\label{eq:K12}
\end{eqnarray}
with
\begin{eqnarray}
F&\equiv F(n,t)= \{c_n^{\dagger}(t), c_0\},\label{eq:defFGbraket}\\
G&\equiv G(n,t)=\{c_n(t), c_0\},  \label{eq:defFGbraket2}\\
Q&\equiv Q(n,t)=\langle c_n^{\dagger}(t) c_0 \rangle_{},\\
W&\equiv W(n,t)=\langle c_n(t) c_0 \rangle_{}.\label{eq:defQWbraket}
\end{eqnarray}
We here used the fact that the anti-commutation relations on the right-hand sides of Eqs.~\eqref{eq:defFGbraket} and \eqref{eq:defFGbraket2} are actually c-numbers; see \ref{sec:AppA}.

For the anisotropic case $\gamma\neq 0$, we obtain the analytic expressions of the functions $F,G,Q$, and $W$ as follows by using the quasiparticle expression \eqref{eq:etapsptp} in the thermodynamic limit:
\begin{eqnarray}
F(n,t)&=\int_{-\pi}^{\pi} \frac{dp}{2\pi} \left(  |s_{p}|^{2}\Phi_{p}(n,t)+ |t_{p}|^{2} \Phi^{*}_{p}(n,t)\right), \label{eq:Fdefint}\\
G(n,t)&=\int_{-\pi}^{\pi} \frac{dp}{2\pi} s_{p}t_{p} \left(\Phi_{p}(n,t)+\Phi^{*}_{p}(n,t)\right),\label{eq:Gdefint}\\
Q(n,t)&=\int_{-\pi}^{\pi} \frac{dp}{2\pi} |t_p|^{2} \Phi^{*}_{p}(n,t),\label{eq:Qdefint}\\
W(n,t)&=\int_{-\pi}^{\pi} \frac{dp}{2\pi} s_{p}t_{p}\Phi^{*}_{p}(n,t), \label{eq:Wdefint}
\end{eqnarray}
where
\begin{eqnarray}
\Phi_{p}(n,t)\equiv \mathrm{e}^{i(\varepsilon(p)t-pn)}.
\end{eqnarray}
For the isotropic case $\gamma=0$, the functions $G$ and $W$ vanish, while
\begin{eqnarray}
F(n,t)&=\int_{-\pi}^{\pi} \frac{dp}{2\pi} \mathrm{e}^{i(\cos{p}+h)t +i pn}  \label{eq:Fdefintxx} \\
&= \mathrm{e}^{i \pi n/2+  iht } J_{n}(t), \label{eq:FGQWfXX}\\
Q(n,t)&= \int_{\{p\,: \, \cos{p}+h\leq0\}} \frac{dp}{2\pi} \mathrm{e}^{i(\cos{p}+h)t+ipn}, \label{eq:Qdefintxx}
\end{eqnarray}
where $J_{n}(t)$ is the Bessel function of the first kind.
We provide an outline of the derivation of these expressions in \ref{sec:AppA}.

The derivation of Eq.~\eqref{eq:DeltaSzForm} can be generalized to other spin-chain Hamiltonians which are mapped into quadratic fermion systems by the Jordan-Wigner transformation as well as for initial states other than the ground state, including a finite-temperature thermal equilibrium state.
For a thermal initial state with the temperature $\beta$, we replace the integrals $ \int_{-\pi}^{\pi}dp$ in Eqs.~\eqref{eq:Qdefint} and \eqref{eq:Wdefint}, and $\int_{\{p\,: \, \cos{p}+h\leq0\}}dp$ in Eq.~\eqref{eq:Qdefintxx} with $ \int_{-\pi}^{\pi}dp \left(1+\exp(\beta \varepsilon(p))\right)^{-1}$.

\subsection{Asymptotic analysis and the velocity of propagation}\label{subsec:asympAnaly}
The propagation velocity of the magnetization change is well characterized by the group velocity of the quasiparticles that are emitted from the impacted site.
From Eqs.~\eqref{eq:DeltaSzForm} and \eqref{eq:Fdefint}--\eqref{eq:Qdefintxx}, we can expect that the dominant component of the wave front propagates with the group velocity at the local extrema.
Here we roughly explain it by approximating the integrals $F,G,Q$, and $W$ in the space-time scaling limit, that is, for a large time $t$ with $v=x/t$ fixed.

The functions in Eqs.~\eqref{eq:Fdefint}--\eqref{eq:Wdefint} have the following integral form in common:
\begin{eqnarray}
I(x,t)\equiv  \int_{-\pi}^{\pi} \frac{dp}{2\pi} g(p)\mathrm{e}^{i(\varepsilon(p)t-px)} ,\label{eq:Ifunc}
\end{eqnarray}
where $g(p)$ is a continuous function for $p\in [-\pi,\pi]$.
Since the magnetization change $\Delta(S^{z}; n, t)$ is expressed by a quadratic sum of the integrals $F,G,Q$, and $W$ in Eqs.~\eqref{eq:Fdefint}--\eqref{eq:Qdefintxx}, we can estimate the behavior of $\Delta(S^{z}; n, t)$ by investigating the asymptotic behavior of Eq.~\eqref{eq:Ifunc}.
For a large $t$ with $v=x/t$ fixed, the leading contribution is obtained from the integral around a stationary point $p^{*}$ at which 
\begin{eqnarray}
\left.\frac{d}{dp}(\varepsilon(p) t -px)\right|_{x=vt}=t\frac{d}{dp}(\varepsilon(p)-vp)=0,
\end{eqnarray}
or $\varepsilon'(p^{*})=v$ holds \cite{wong2001asymptotic}.
Then we  expand $\varepsilon(p)$ around $p^{*}$ as
\begin{eqnarray}
\varepsilon(p)\sim\varepsilon(p^{*}) + v(p-p^{*}) +\frac1{\kappa!}\varepsilon^{(\kappa)}(p^{*})(p-p^{*})^{\kappa},
\end{eqnarray} 
where
\begin{eqnarray}
\kappa\equiv \min_{m\geq2}\, \left\{ m \left| \, \frac{d^{m}}{dp^{m}}\varepsilon(p^{*})=\varepsilon^{(m)}(p^{*}) \neq 0\right.\right\},\label{eq:kappa}
\end{eqnarray}
and we assumed $g(p^{*})\neq0$ and $v\leq \max_{p}v_{\rm g}(p)$ so that the stationary point $p^{*}$ may exist.
We perform the Fresnel integral to obtain 
\begin{eqnarray}
\fl
I(vt,t) \simeq  \int_{-\pi}^{\pi} \frac{dp}{2\pi}g(p^{*})\exp\left[{it\left(\varepsilon(p^{*}) + \varepsilon'(p^{*})(p-p^{*}) +\frac1{\kappa!}\varepsilon^{(\kappa)}(p^{*})(p-p^{*})^{\kappa} \right)-ipvt}\right]\nonumber \\
 \simeq  g(p^{*})\mathrm{e}^{it\varepsilon(p^{*})}\int_{-\pi}^{\pi} \frac{dp}{2\pi} \exp\left[{it\frac1{\kappa!}\varepsilon^{(\kappa)}(p^{*})p^{\kappa} }\right] = O\left(t^{-1/\kappa}\right).\label{eq:Ifuncstation}
\end{eqnarray}
(We present more precise approximations in Sec.~\ref{sec:AiryNew}.)

This shows that the integral \eqref{eq:Ifuncstation} generally decays as $t^{-1/2}$ except that it decays slower than $t^{-1/2}$ when we choose $v$ to be the group velocity $v_{\rm g}(p)$ at one of its local extrema, where the corresponding stationary point $p^{*}$ satisfies $\varepsilon''(p^{*})=0$, and thereby $\kappa\geq3$.
Therefore the integral \eqref{eq:Ifunc} yields wave fronts which propagate with the group velocity at its local extrema, forming the profile of a light cone and standing out from the bulk inside the light cone.

For the anisotropic case $\gamma\neq0$, the dispersion relation \eqref{eq:varEpsilon} can have two inflection points in $0<p\leq \pi$ for some parameter regions, which means that $v_{\rm g}(p)$ can have two local extrema in $0<p\leq \pi$.
(We only describe the inflection points in $0<p\leq \pi$ hereafter since the dispersions \eqref{eq:varEpsilon} and \eqref{eq:DispXX} are even functions of $p$.)
In this case, there generally appear two wave fronts propagating with the velocities $V_{1}\equiv |v_{\rm g}(p^{*})|$ and $V_{2}\equiv |v_{\rm g}(p^{**})|$, where $p^{*}$ and $p^{**}$ denote the inflection points as in $\varepsilon''(p^{*})=\varepsilon''(p^{**})=0$, and we assumed $V_{1}\geq V_{2}$ without loss of generality.
The second velocity $V_{2}$ is defined only when the dispersion relation $\varepsilon(p)$ has two inflection points in $0<p\leq\pi$.

\section{Light-cone dynamics in various phases}\label{sec:res}
In this section, we calculate the magnetization change \eqref{eq:DeltaSzForm} by numerical integration of Eqs.~\eqref{eq:Fdefint}--\eqref{eq:Qdefint} and \eqref{eq:Qdefintxx}, and investigate the propagation dynamics under the local-impact protocol analytically.
For the model parameters, we mainly investigate the region $0\leq h \leq2$ and $0\leq\gamma\leq1$.
We particularly present the results for the local impact $U_{\rm loc}=\mathrm{e}^{i\theta S_{0}^{z}}$ with $\theta=2\pi/3$. The choice of $\theta$ makes only subtle change in the propagation dynamics because quasiparticles with any $p$ are excited anyway as we stressed at the end of Sec.~\ref{subsec:digonal}.

We observe that the local impact creates a ballistically propagating wave fronts, forming a light cone, except for the case of $\gamma=0$, $h\geq1$, in which no dynamics is obtained since the ground state becomes an eigenstate of the local impact $U_{\rm loc}=\mathrm{e}^{i\theta S_{0}^{z}}$ as well as $S_{0}^{z}$ (see the expectation value of $S_{0}^{z}$ for $\gamma=0$ in Fig.~\ref{fig:SzGsVaryh}), and for the case of $\gamma=1$ and $h=0$, namely when the model reduces to the trivial Ising model $H=\sum_{n=0}^{L-1}S_{n}^{x}S_{n+1}^{x}$, in which the local impact only causes an oscillation in $S^{z}_{0}$ and does not spatially propagate, {\it i.e. $\Delta(S^{z}; n\neq0, t)=0$}.
We do not consider these exceptional cases hereafter.

First, we provide a phase diagram according to the number of inflection points, which is relevant in investigating the propagation of quasiparticles, and then present some results obtained by numerically integrating the functions \eqref{eq:Fdefint}--\eqref{eq:Wdefint} with particular interest in the speed of the propagating wave fronts.
After that we show that the wave front decays in a power law in time with exponents depending on where the model is located in the phase diagram.

\subsection{Phase diagram and the propagation dynamics}
\begin{figure}
\begin{indented}\item[]
\includegraphics[width=0.7\textwidth]{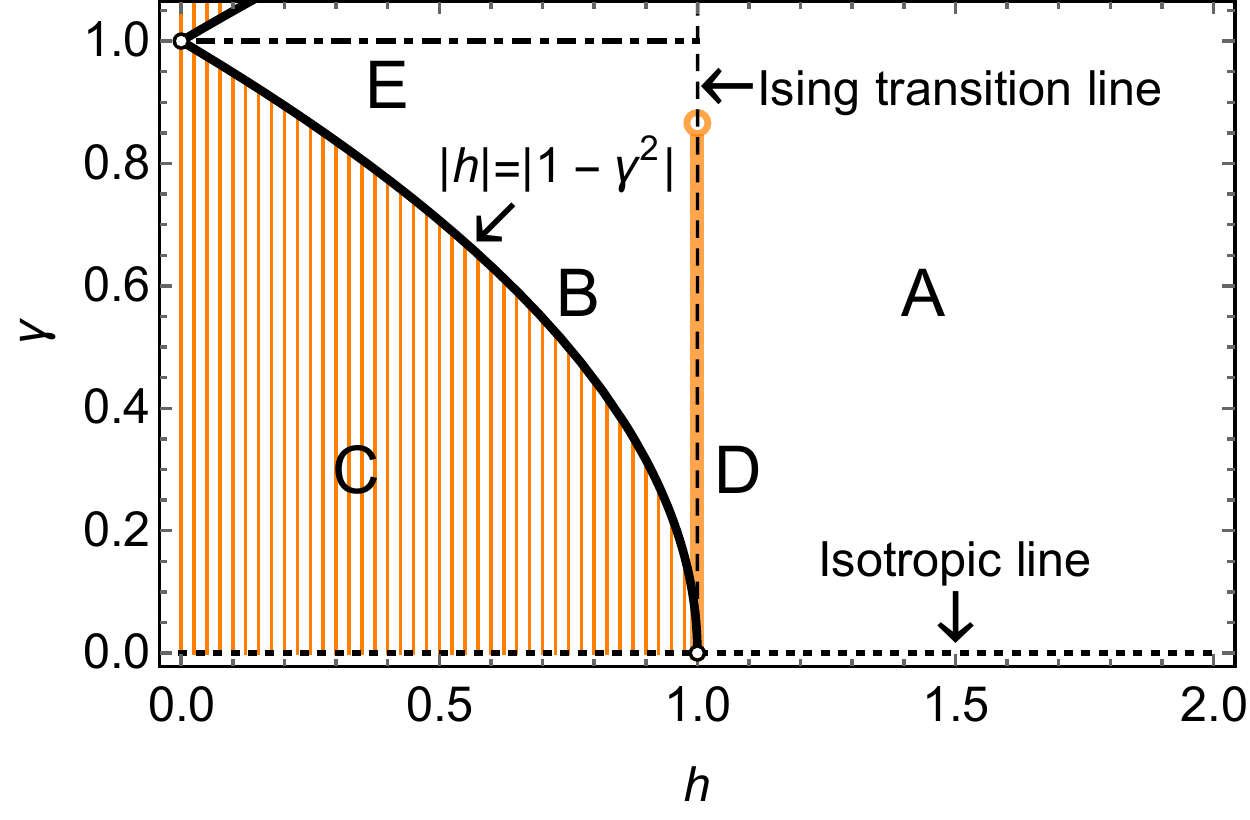}
\caption{The phase diagram according to the number of inflection points of the dispersion relation \eqref{eq:varEpsilon} in the parameter region $0 \leq h \leq2$, $0<\gamma\leq1$.
The region $A$ represents the parameter region $|h|>|1-\gamma^{2}|$, excluding the orange thick line for $h=1$, on which $\varepsilon(p)$ has only one inflection point in $0<p\leq\pi$.
The regions $B$, $C$, and $D$ represent the parameter regions $|h|=|1-\gamma^{2}|$ (the black thick carve), $|h|<|1-\gamma^{2}|$ (the region with orange vertical lines), and $h=1$ with $0<\gamma<\sqrt{3}/2=\gamma_{c}$ (the orange thick line), respectively.
In $B$, $C$, and $D$, the dispersion $\varepsilon(p)$ has two inflection points in $0<p\leq\pi$.
On the chain line $E$ (the region $\gamma=1$ with $0<h<1$) and on the Ising transition line $h=1$ (the broken line), the relation $\cos{p^{*}}+h=0$ holds for one of the inflection point $p^{*}$ in $0<p\leq\pi$ (see Sec.~\ref{subsec:AiryFail}).
On the isotropic line $\gamma=0$ (the dotted line), the dispersion relation is given by Eq.~\eqref{eq:DispXX}, and its inflection points are $p=\pm\pi/2$.
}
 \label{fig:phasedig}
\end{indented}
\end{figure}

\begin{figure}
\begin{indented}\item[]
\centering
\subfigure[]{
\centering
\includegraphics[width=0.4\textwidth]{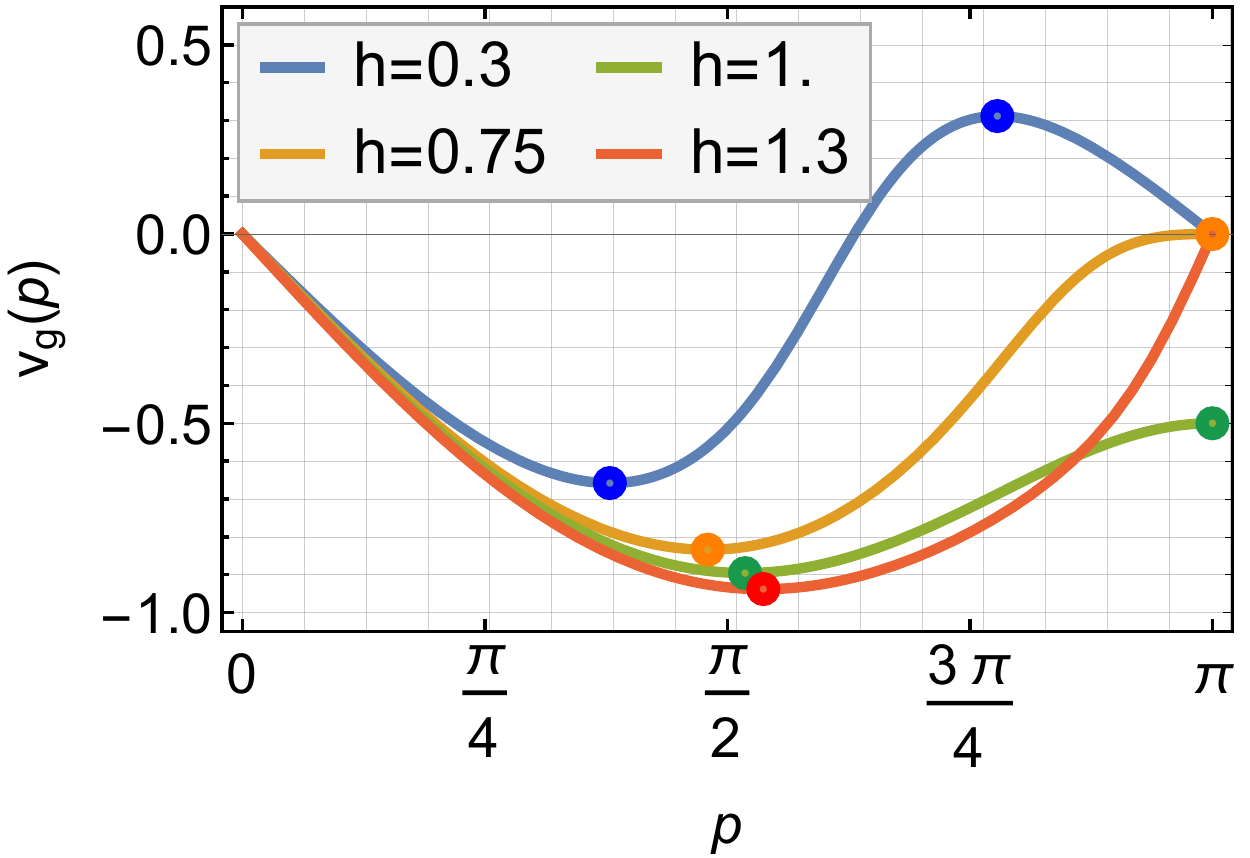}
 \label{fig:DispersionVgG05}
 }
\subfigure[]{
\centering
 \includegraphics[width=0.4\textwidth]{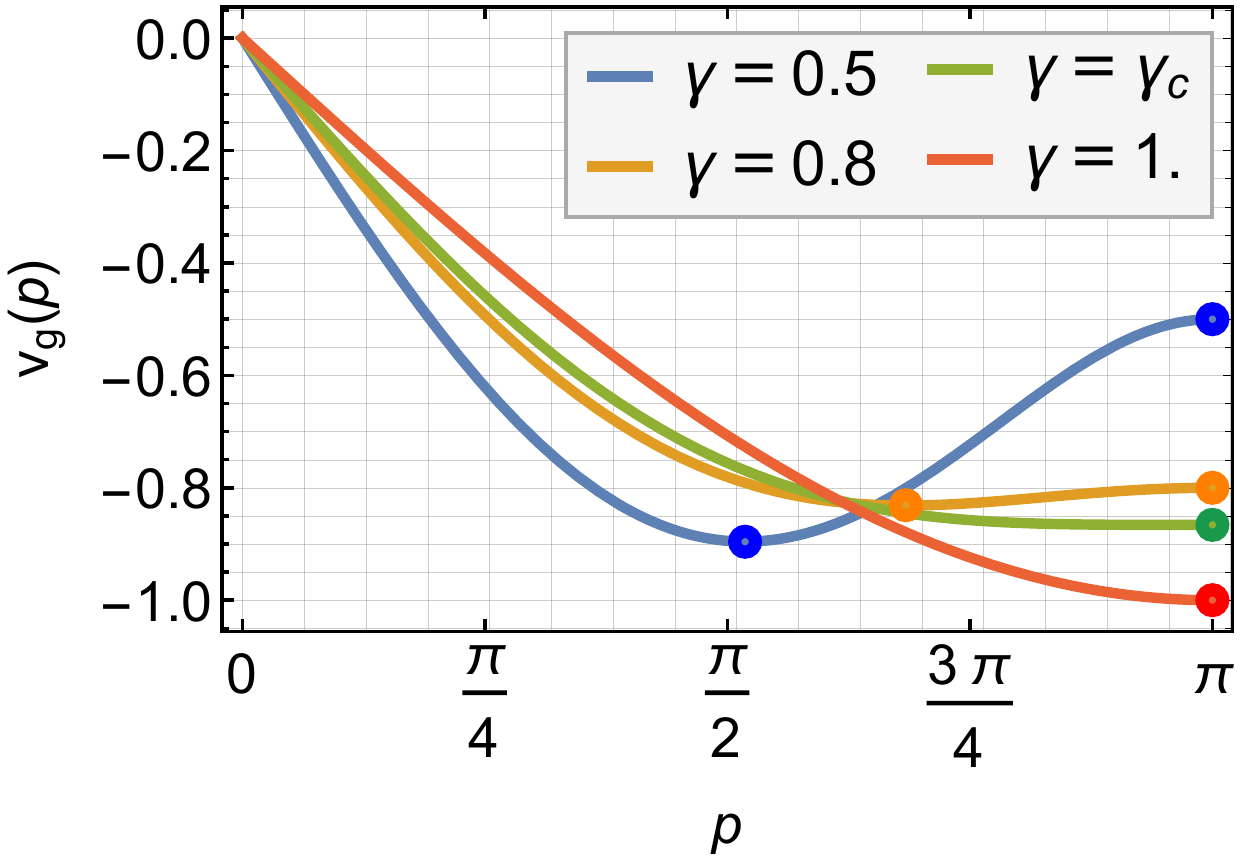}
  \label{fig:DispersionVgH1}
}
 \caption{The group velocity $v_{\rm g}(p)=\varepsilon'(p)$ for the $XY$ model.
Small circles indicate the local extrema of the group velocity for each model-parameter set.
(a) The group velocity in the case of $\gamma=0.5$ with various values of the magnetic field $h=0.3$ (region $C$ in Fig.~\ref{fig:phasedig}), $h=0.75$ (region $B$), $h=1$ (region $D$) and $h=1.3$ (region $A$).
(b) The group velocity in the case of $h=1$ (the Ising transition line in Fig.~\ref{fig:phasedig}) with various values of the anisotropy $\gamma=0.5$, $0.8$ (region $D$), $1$ and $\gamma=\gamma_{c}\sim0.866$ (region $A$).
In both panels we plot functions only for $0\leq p \leq \pi$ since $\varepsilon(p)$ is an even function of $p$.}
\label{fig:DispersionH1bandd}
\end{indented}
\end{figure}

As we have explained in Sec.~\ref{subsec:asympAnaly}, the number of inflection points of the dispersion relation generally corresponds to the number of propagating wave fronts.
First we show in Fig.~\ref{fig:phasedig} the phase diagram according to the number of inflection points in $0<p\leq\pi$.
We also provide in Fig.~\ref{fig:DispersionH1bandd} plots of the group velocities $v_{\rm g}(p)=\varepsilon'(p)$ for some parameter sets in the regions in the phase diagram.
Note that the inflection points of the dispersion relation correspond to the local extrema of the group velocity.
For $|h|\leq|1-\gamma^{2}|$ with $\gamma\neq0$ (the regions $B$ and $C$ in Fig.~\ref{fig:phasedig}), and for $h=1$ with $0<\gamma<\sqrt{3}/2=\gamma_{c}$ (the region $D$ in Fig.~\ref{fig:phasedig}), the dispersion $\varepsilon(p)$ has two inflection points in $0<p\leq\pi$, whereas in the other cases it has only one inflection point.

\begin{figure}
\begin{indented}\item[]
\subfigure[]{
\centering
\includegraphics[width=0.33\textwidth]{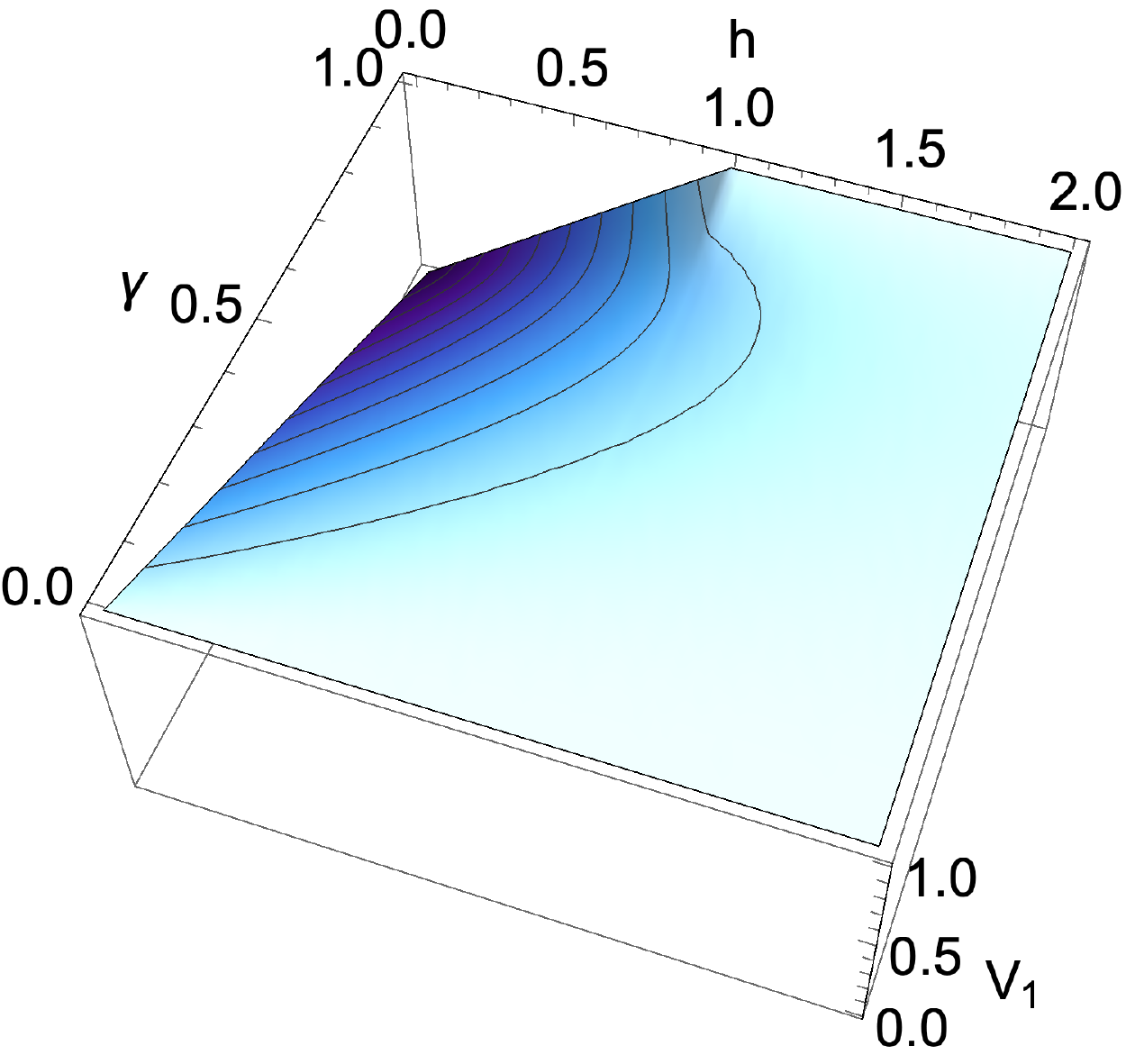}
\label{fig:Ep_Max_Velo(527)}
\includegraphics[width=0.045\textwidth]{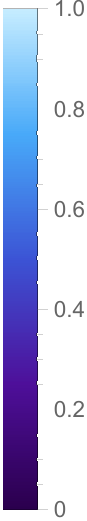}
}
\subfigure[]{
\includegraphics[width=0.33\textwidth]{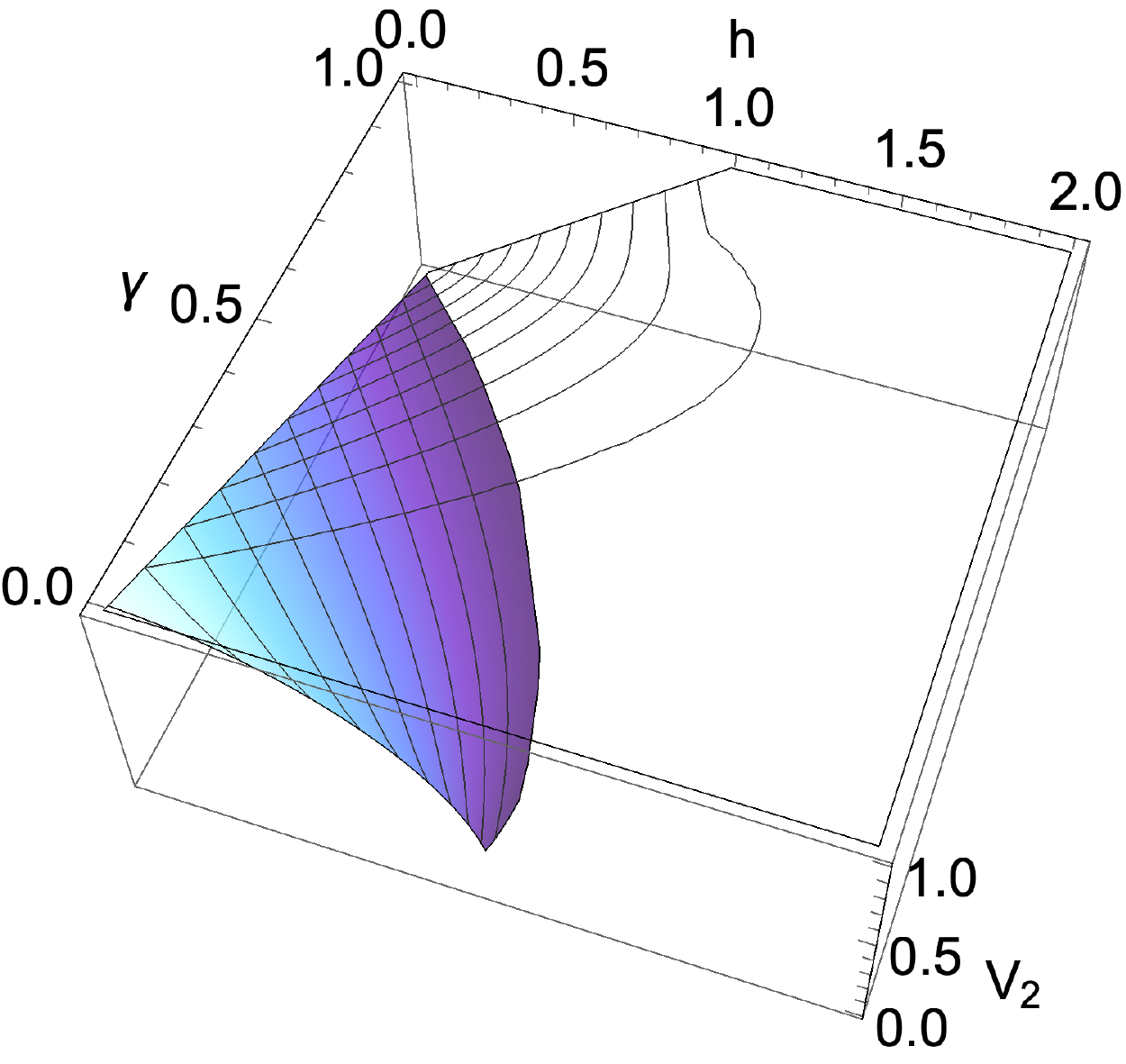}
\label{fig:Ep_Max_Velo_double(527)}
\includegraphics[width=0.045\textwidth]{fig19.pdf}
}
\caption{(a) The maximum group velocity $V_{1}$ of the $XY$ model with the anisotropy $\gamma$ in a magnetic field $h$.
(b) The group velocity for the second local maximum $V_{2}$, which is defined only for $|h|\leq |1-\gamma^{2}|$ with $\gamma\neq0$, as well as for $h=1$ with $0<\gamma<\gamma_{c}$. 
The velocity  $V_{2}$ in the latter case is not shown in (b), where $V_{2}=|v_{\rm g}(\pi)|=\gamma$.
}\label{fig:VeloAnaly}
\end{indented}
\end{figure}
\begin{figure*}
\centering
\subfigure[]{
\centering
\includegraphics[width=0.31\textwidth]{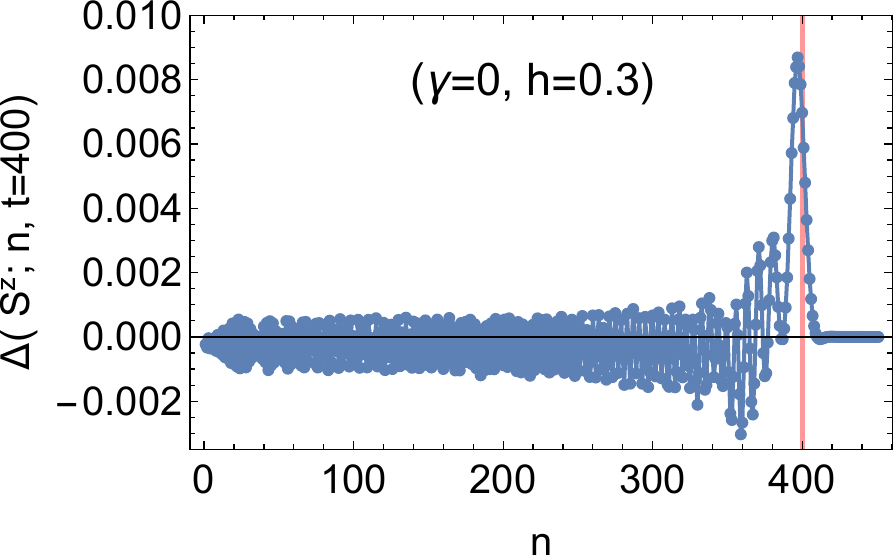}
\label{fig:ProFileG0H03}
}
\subfigure[]{
\centering
\includegraphics[width=0.31\textwidth]{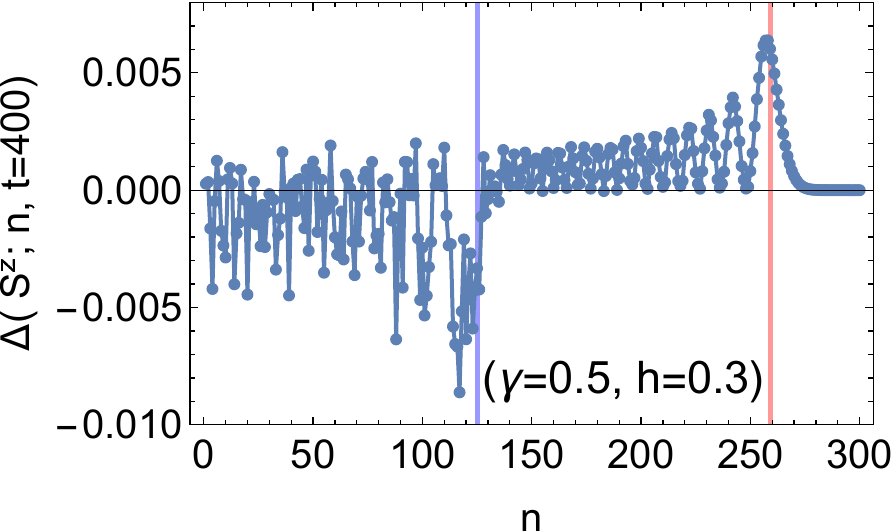}
\label{fig:ProFileG05H03}
}
\centering
\subfigure[]{
\centering
\includegraphics[width=0.31\textwidth]{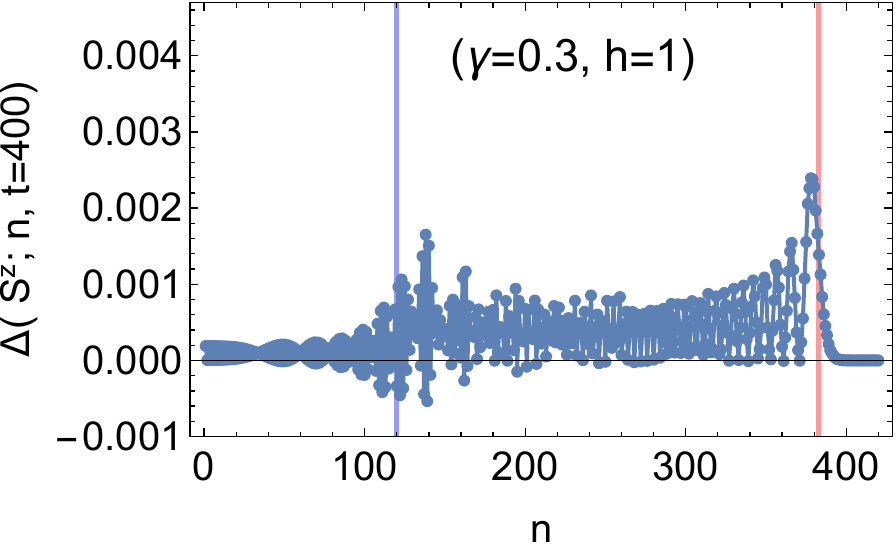}
\label{fig:ProFileG03H1}
}
\subfigure[]{
\centering
\includegraphics[width=0.31\textwidth]{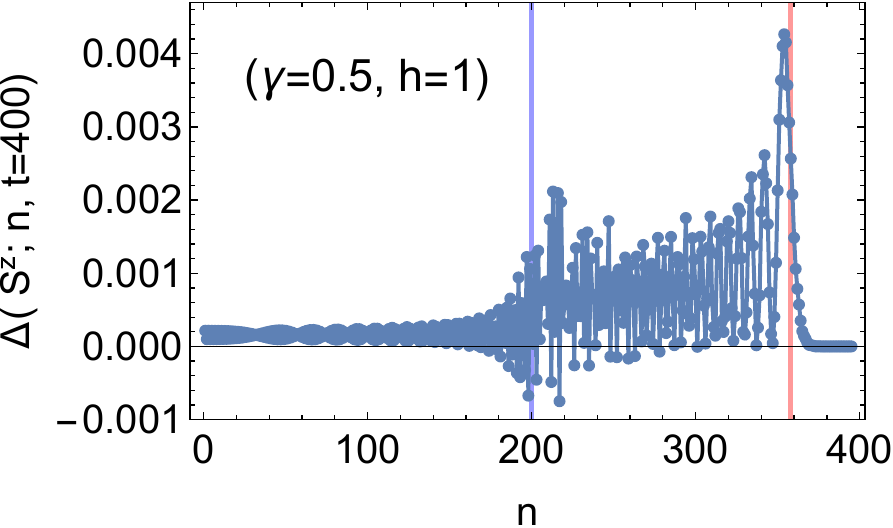}
\label{fig:ProFileG05H1}
}
\subfigure[]{
\centering
\includegraphics[width=0.31\textwidth]{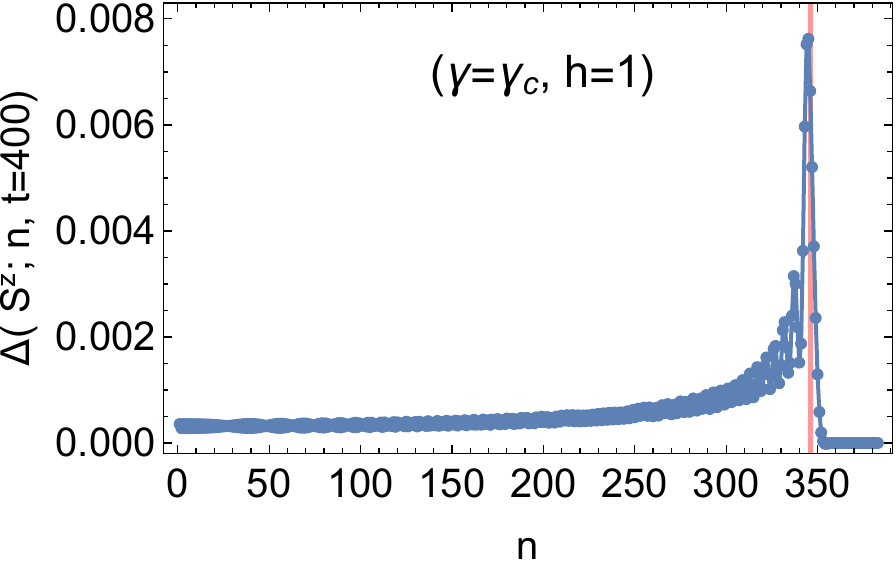}
\label{fig:ProFileGCRIH1}
}
\subfigure[]{
\centering
\includegraphics[width=0.31\textwidth]{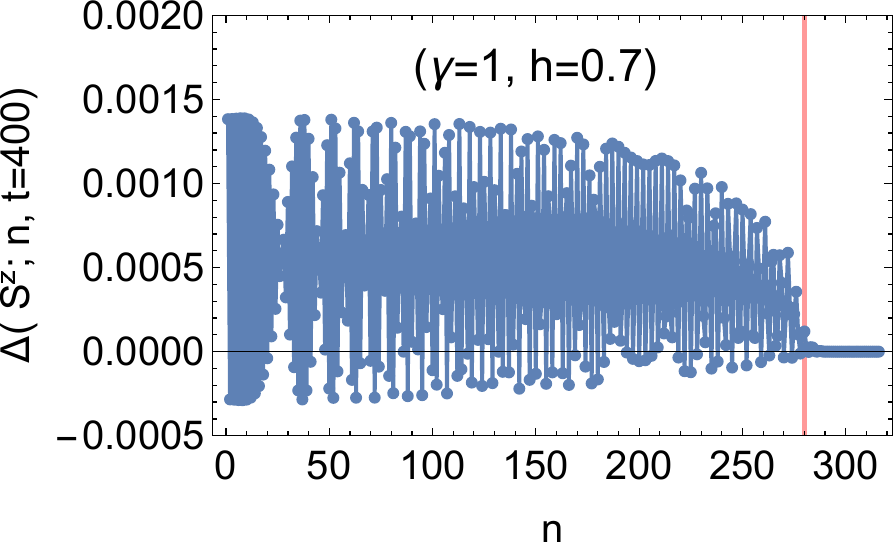}
\label{fig:ProFileG1H07}
}
\subfigure[]{
\centering
\includegraphics[width=0.31\textwidth]{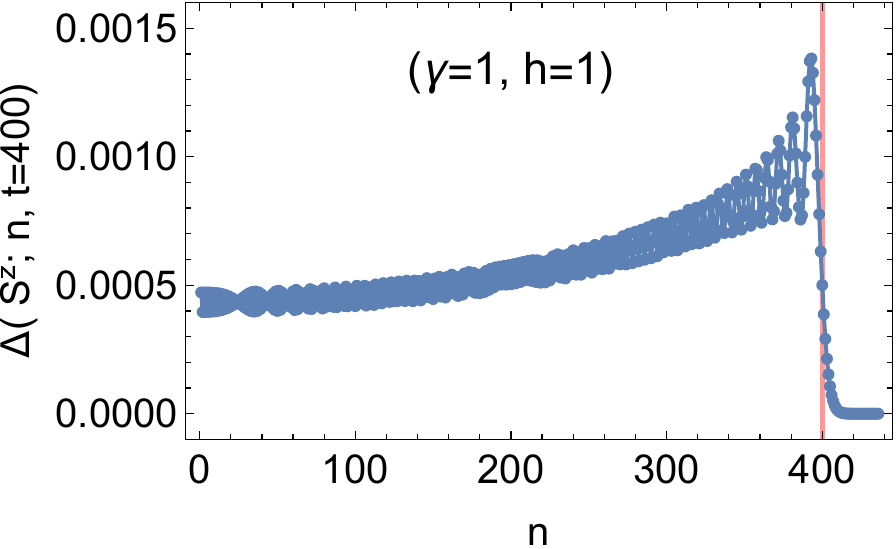}
\label{fig:ProFileG1H1}
}
\subfigure[]{
\centering
\includegraphics[width=0.31\textwidth]{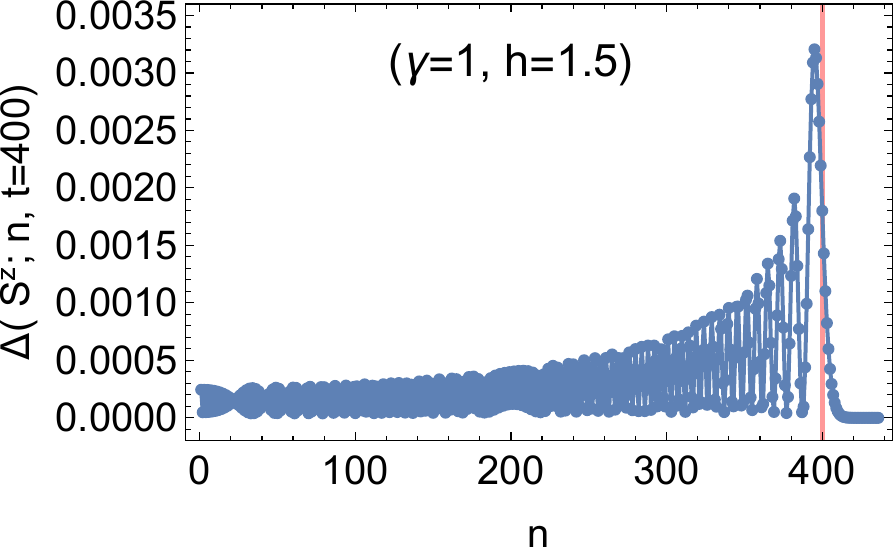}
\label{fig:ProFileG1H15}
}
\caption{Numerical calculation of the magnetization change $\Delta(S^{z}; n, t)$ at time $t=400$ for eight parameter sets of the $XY$ model.
The vertical lines indicate the points $n=V_{1}t$ (red lines) and $n=V_{2}t$ (blue lines).
The values of the model parameters, {\it i.e.}, the anisotropy $\gamma$ and the magnetic field $h$ are shown in each panel.
The location of the parameter sets in Fig.~\ref{fig:phasedig} are: (a) on the isotropic line; (b) region $C$; (c), (d) region $D$; (e) upper edge of the line $D$ (namely region $A$); (f) region $E$; (g) the right edge of the line $E$ (namely region $A$); (h) region $A$.
}
\label{fig:Profile8}
\end{figure*}

The maximum group velocity $V_{1}$ and the group velocity at its second local maximum $V_{2}$ are shown in Fig.~\ref{fig:VeloAnaly}.
As we will observe below, they mostly give good estimates of the location of the wave fronts.
We obtained the velocities by numerically searching the inflection points of $\varepsilon(p)$ in the parameter region $0.05\leq\gamma\leq1$, $0\leq h \leq2$.
In the isotropic case $\gamma=0$, namely on the isotropic line in Fig.~\ref{fig:phasedig}, the maximum group velocity is always unity and there are no second local maxima as in Fig.~\ref{fig:ProFileG0H03} because the dispersion relation \eqref{eq:DispXX} is $\varepsilon(p)=\cos{p}+h$.
For $h=0$ with $0<\gamma<1$, the second velocity $V_{2}$ coincides with the first one $V_{1}$ and hence there appears only one wave front even though the number of inflection points in $0<p\leq \pi$ is two.

Incidentally, we show in \ref{sec:AppLR} that the Lieb-Robinson velocity is much faster than $V_{1}$ and $V_{2}$. The parameter dependence is also essentially different from the one in Fig.~\ref{fig:VeloAnaly}.

Now we provide results of the dynamics of the magnetization change $\Delta(S^{z}; n, t)$ under the local-impact protocol.
We obtained the dynamics of $\Delta(S^{z}; n, t)$ in the thermodynamic limit by numerical integration of Eqs.~\eqref{eq:Fdefint}--\eqref{eq:Qdefintxx}.
\begin{figure}
\begin{indented}\item[]
\subfigure[]{
\centering
\includegraphics[width=0.4\textwidth]{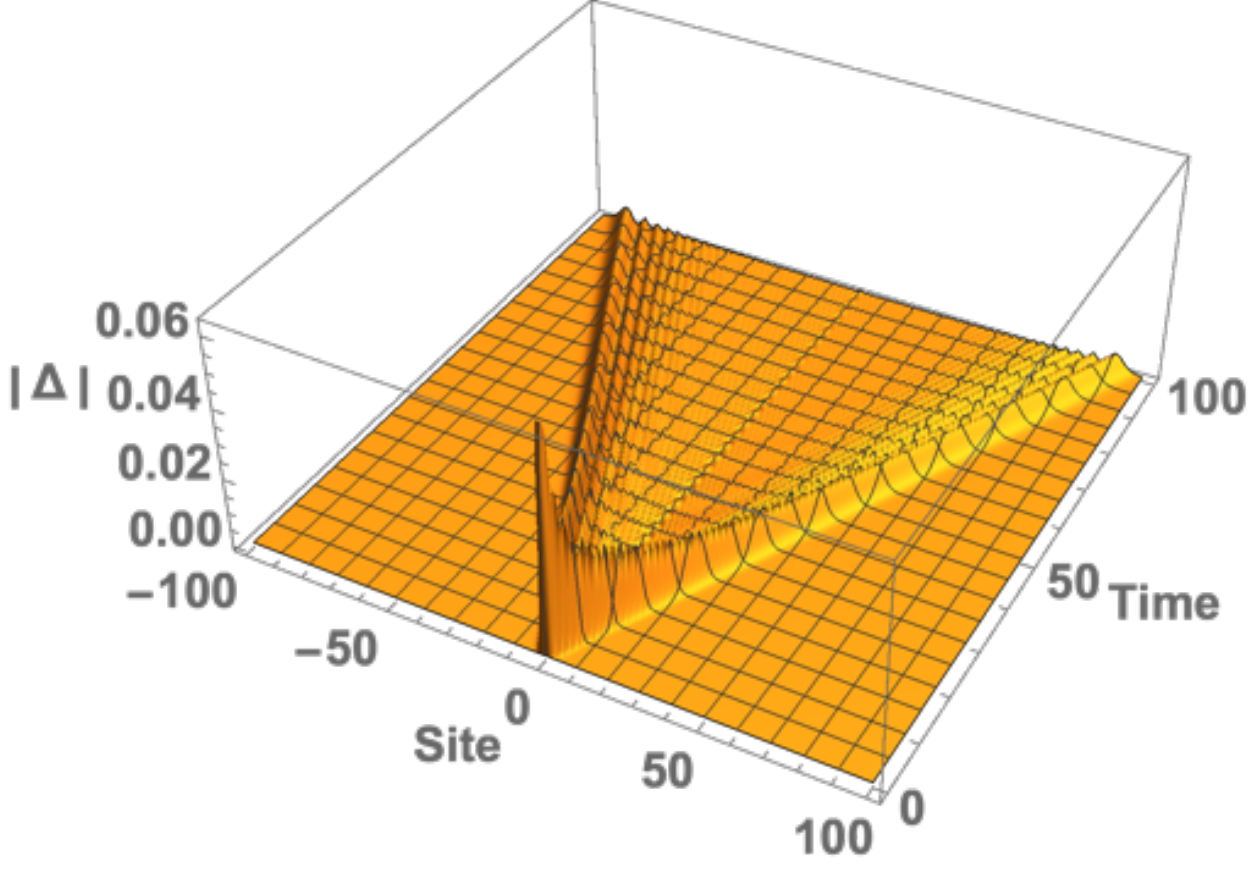}
\label{fig:3DC}
}
\subfigure[]{
\centering
\includegraphics[width=0.4\textwidth]{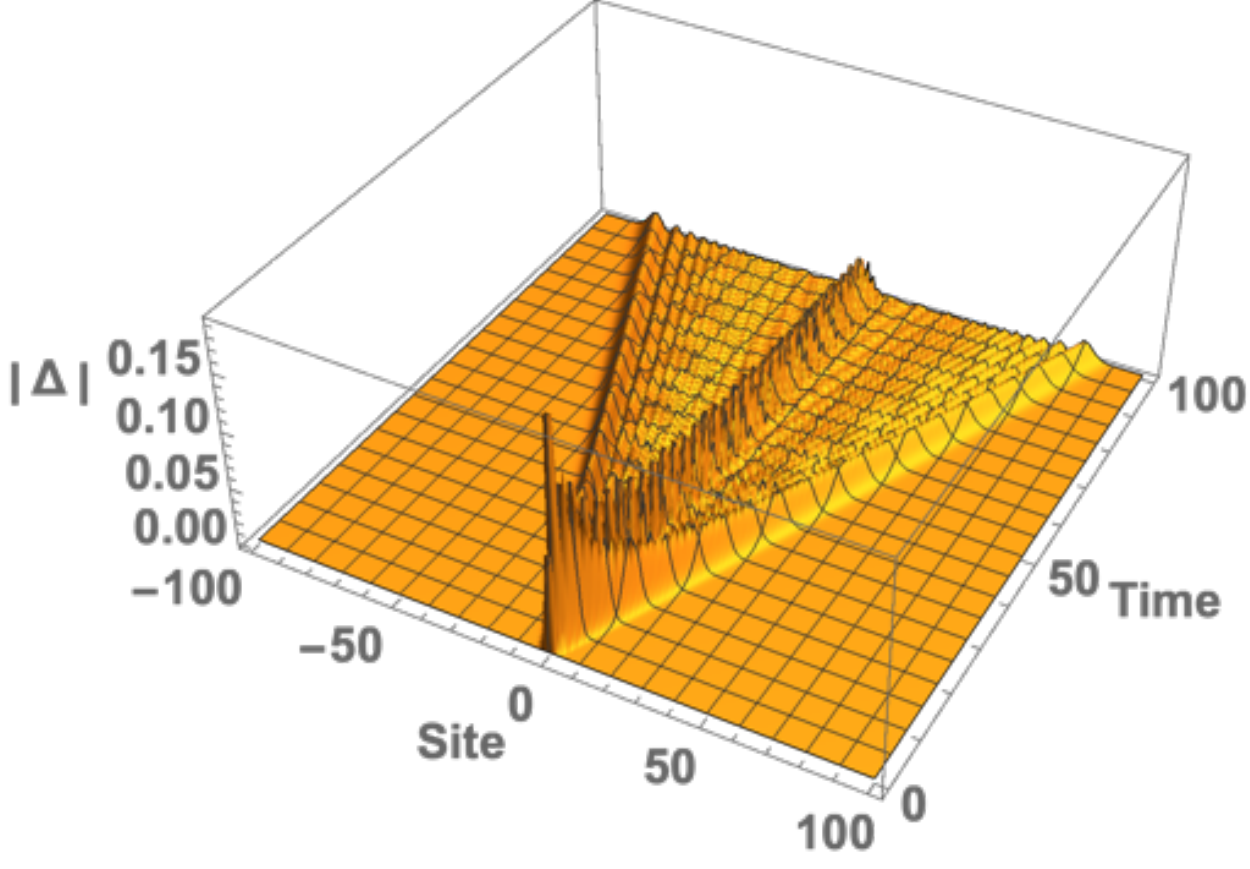}
\label{fig:3DB}
}
\subfigure[]{
\centering
\includegraphics[width=0.4\textwidth]{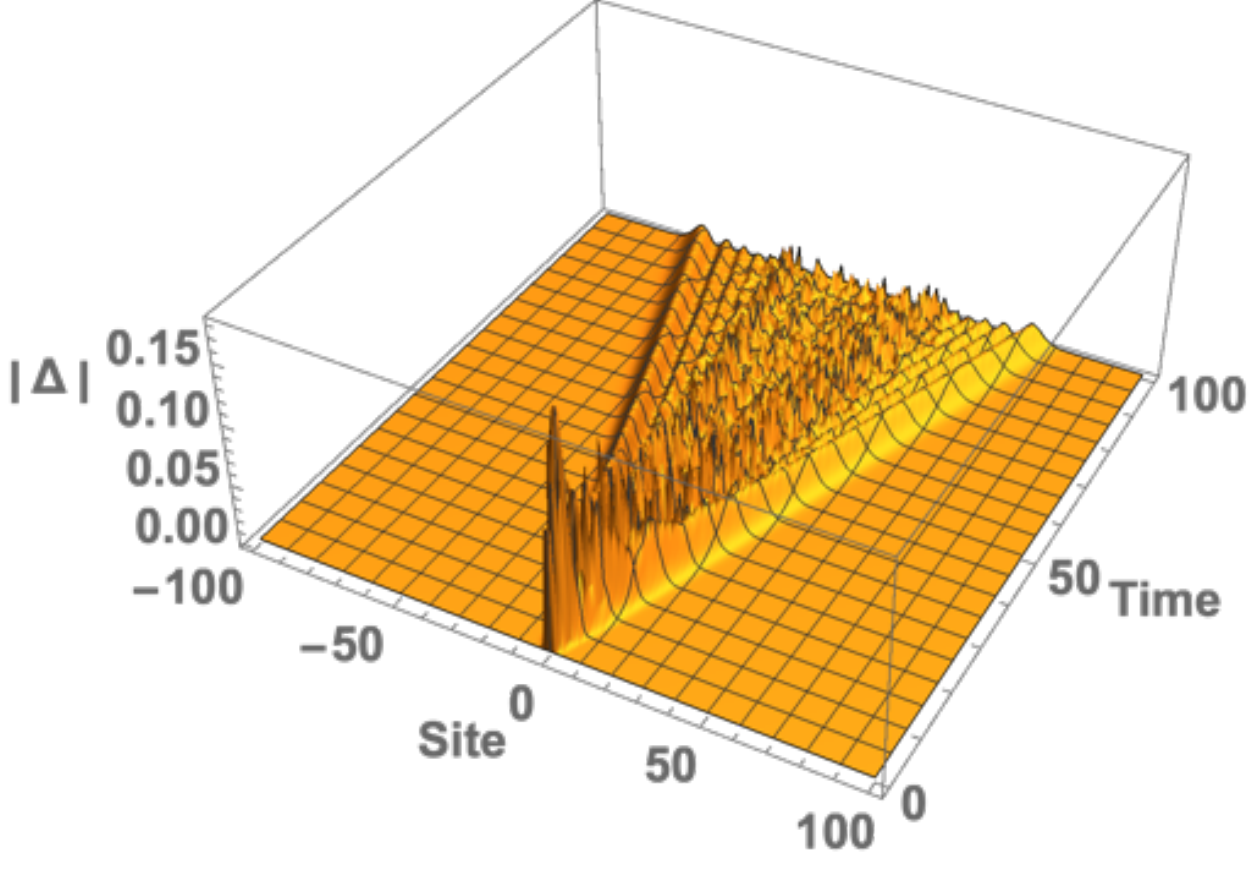}
\label{fig:3DA}
}
\subfigure[]{
\centering
\includegraphics[width=0.4\textwidth]{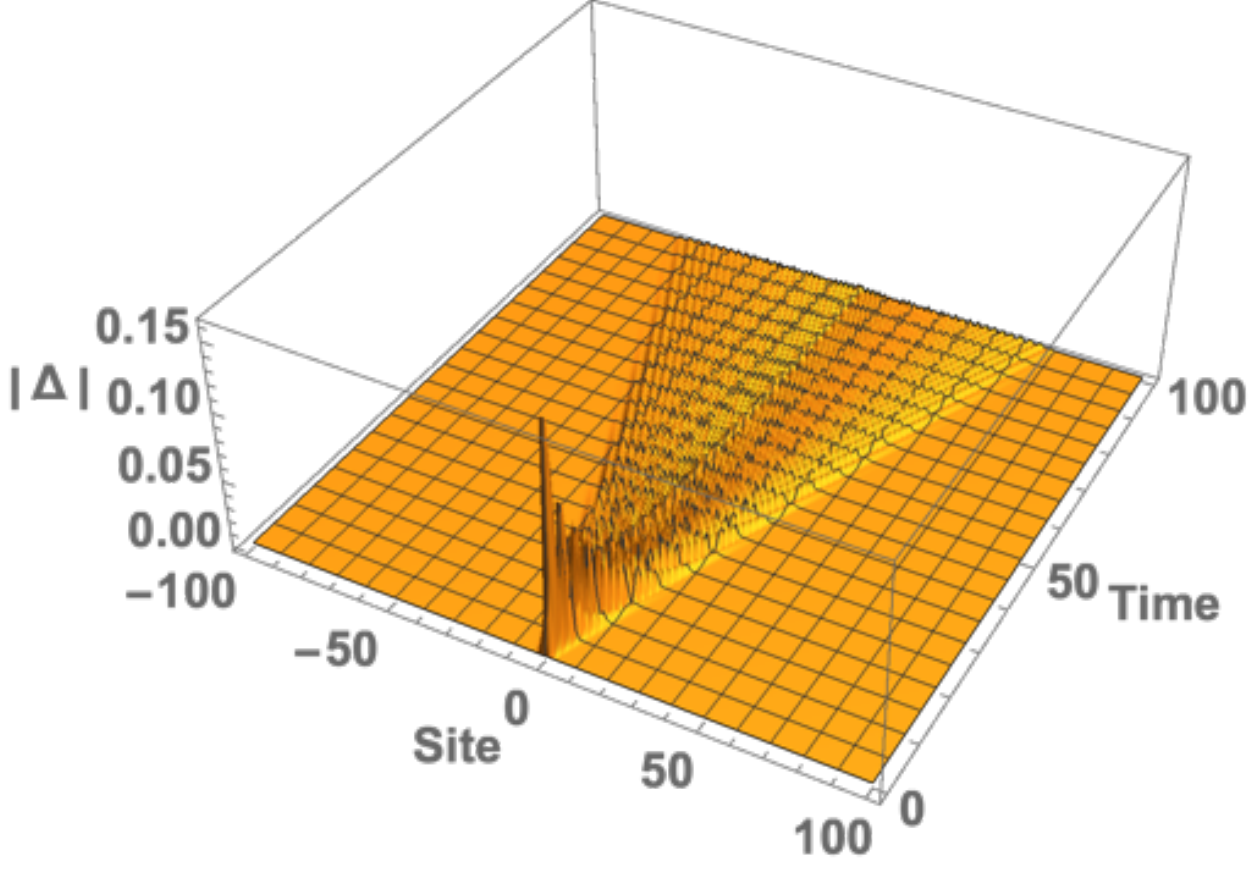}
\label{fig:3DIsing}
}
\caption{Numerical calculation of dynamics of $|\Delta|:=|\Delta(S^{z}; n, t)|$ for four parameter sets of the $XY$ model.
(a) The model parameters are $\gamma=0.5$, $h=1.3$ (region $A$ in Fig.~\ref{fig:phasedig}), and the group velocities at the local extrema, which we defined in Sec.~\ref{subsec:SettingLast}, are estimated at $V_{1}\simeq0.93$.
(b) The model parameters are $\gamma=0.5$, $h=0.75$ (region $B$), and $V_{1}\simeq0.83$ and $V_{2}=0$.
(c) The model parameters are $\gamma=0.5$, $h=0.3$ (region $C$), and  $V_{1}\simeq0.66$ and $V_{2}\simeq0.31$.
(d) The model parameters are $\gamma=1$, $h=0.7$ (region $E$), and $V_{1}=0.7$.
}
\label{fig:3ds}
\end{indented}
\end{figure}

Figure~\ref{fig:3ds} shows the dynamics of $|\Delta(S^{z}; n, t)|$ for four sets of the model parameters in the regions of $A$, $B$, $C$, and $E$ in the phase diagram.
We also provide the profiles of the magnetization change $\Delta(S^{z}; n, t)$ at time $t=400$ in Fig.~\ref{fig:Profile8} for eight parameter sets.

In most cases, a pair of wave fronts propagates ballistically with a clear peak, forming a light cone, as is exemplified in Figs.~\ref{fig:3DC}--\ref{fig:3DA}, and the magnetization change is exponentially suppressed at the sites $n$ outside of the light cone.
All panels of Fig.~\ref{fig:Profile8} except for Fig.~\ref{fig:ProFileG1H07} demonstrate that the first peak position agrees well with the red vertical line, which indicates $V_{1}t$ obtained from the analysis in Sec.~\ref{subsec:asympAnaly}.

In addition, there emerges another pair of wave fronts inside the light cone when the parameter set is located in the region $C$ in the phase diagram, owing to the second local extrema of the group velocity of the quasiparticles; see Fig.~\ref{fig:3DA}.
Figure~\ref{fig:ProFileG05H03} demonstrates that the position of the second peak also agrees well with the blue vertical line, which indicates $V_{2}t$.
These observations confirm the validity of our analysis in Sec.~\ref{subsec:asympAnaly}.

The existence of the second wave front in the $XY$ model has been suggested in some studies \cite{RoschmidtechoXY, XYFagotti2008OSRE}.
In the global-quench protocol, where the quasiparticle picture \cite{calabrese2005evolution,Calabrese_2006, Alba7947} has been used to describe the information propagation dynamics in integrable systems, the second wave front would be blurred by all waves from other points on a chain.
Nevertheless, Ref.~\cite{XYFagotti2008OSRE} numerically observed that quasiparticles with the mode at the second local extremum of $v_{\rm q}(p)$ can carry a dominant part of information which contributes to the entanglement growth in a global-quench setting.

On the line $B$, the second set of wave fronts on the right and left sides merge in the middle to create a ridge at $n=0$ as in Fig.~\ref{fig:3DB}, which we refer to as a ``frozen'' wave front.
This is consistent with the fact that $V_{2}=0$ on the line $B$.
We show below in Sec.~\ref{sec:analyRidge} that the frozen wave front decays slower than the first wave front, as we can observe in Fig.~\ref{fig:3DB}.

When the parameter set is located on the line $E$ in Fig.~\ref{fig:phasedig}, where the $XY$ model reduces to the transverse Ising model with the magnetic filed $0<h<1$, we observe no clear peak around the wave front, as is shown in Fig.~\ref{fig:3DIsing} and Fig.~\ref{fig:ProFileG1H07}, whereas a peak appears around the wave front for $\gamma=1$ with $h\geq1$ as shown in Figs.~\ref{fig:ProFileG1H1} and \ref{fig:ProFileG1H15}.
We will reconsider the behavior in Fig.~\ref{fig:ProFileG1H07} below in Sec.~\ref{subsec:AiryFail}.

On the line $D$ in Fig.~\ref{fig:phasedig}, a second local extremum of $v_{\rm g}(p)$ emerges, and hence we would expect the appearance of a second wave front as is the case of the region $C$, but it is in fact hard to identify it in Figs.~\ref{fig:ProFileG03H1} and \ref{fig:ProFileG05H1}.
In Sec.~\ref{subsec:AiryFail} we discuss the origin of this behavior analytically.
The second extremum disappears at the upper end of the line $D$, and hence we obtain a single wave front in Fig.~\ref{fig:ProFileGCRIH1}.

\subsection{Asymptotic behavior of the wave fronts}\label{sec:decay}\label{sec:LongTimeProfile}
\begin{figure}
\centering
\subfigure[]{
\centering
\includegraphics[width=0.31\textwidth]{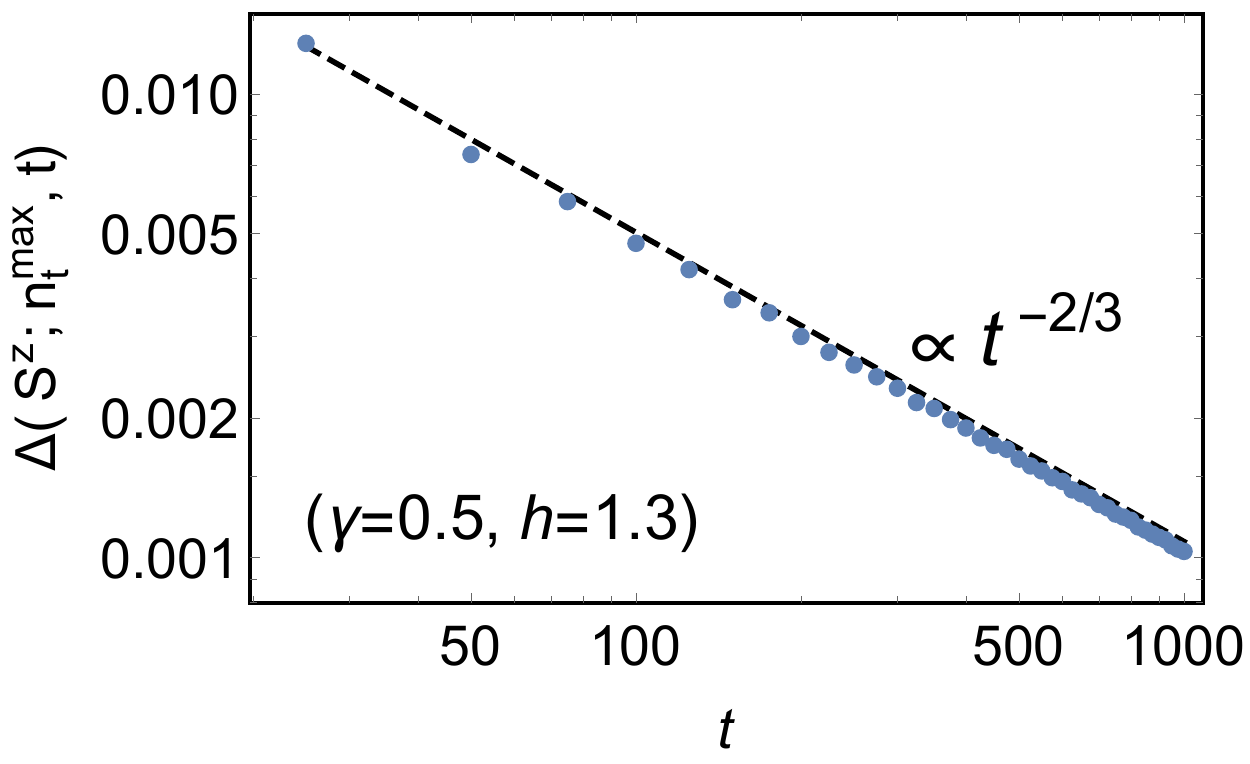}
\label{fig:waveDecayC}
}
\subfigure[]{
\centering
\includegraphics[width=0.31\textwidth]{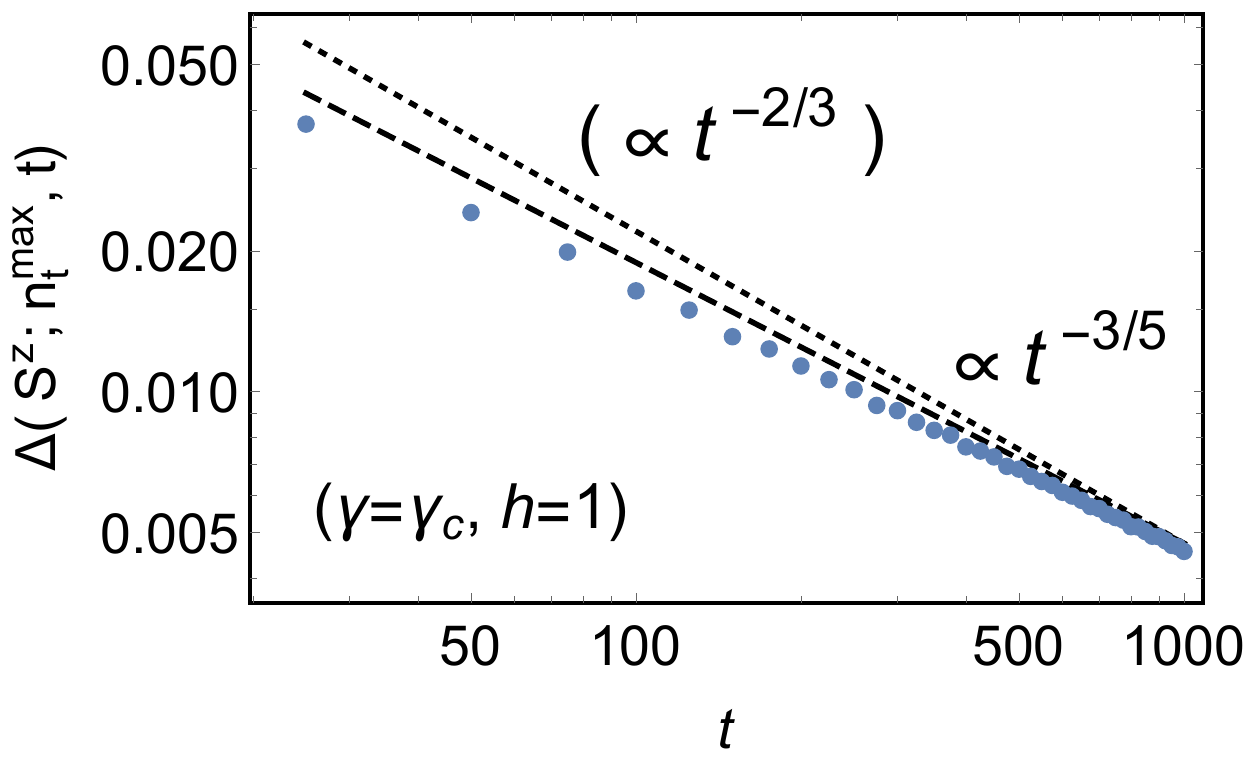}
\label{fig:waveDecayH1Gcri}
}
\subfigure[]{
\centering
\includegraphics[width=0.31\textwidth]{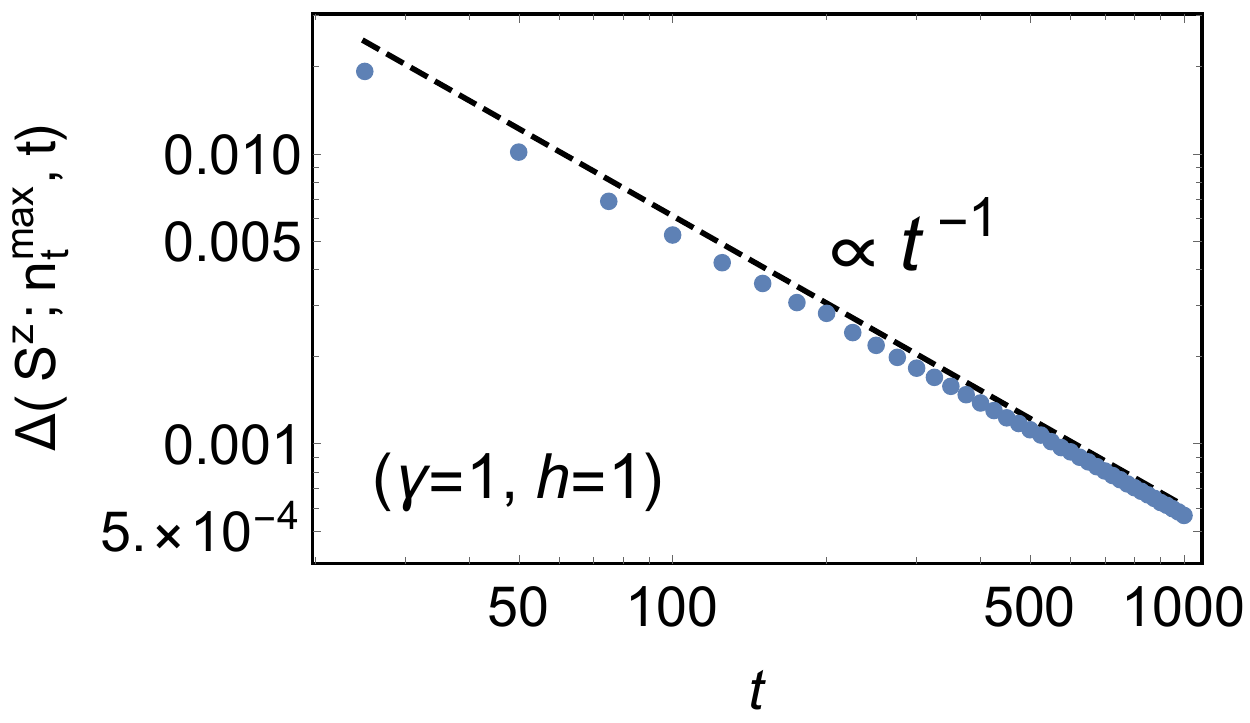}
\label{fig:waveDecayH1G1}
}
\subfigure[]{
\centering
\includegraphics[width=0.31\textwidth]{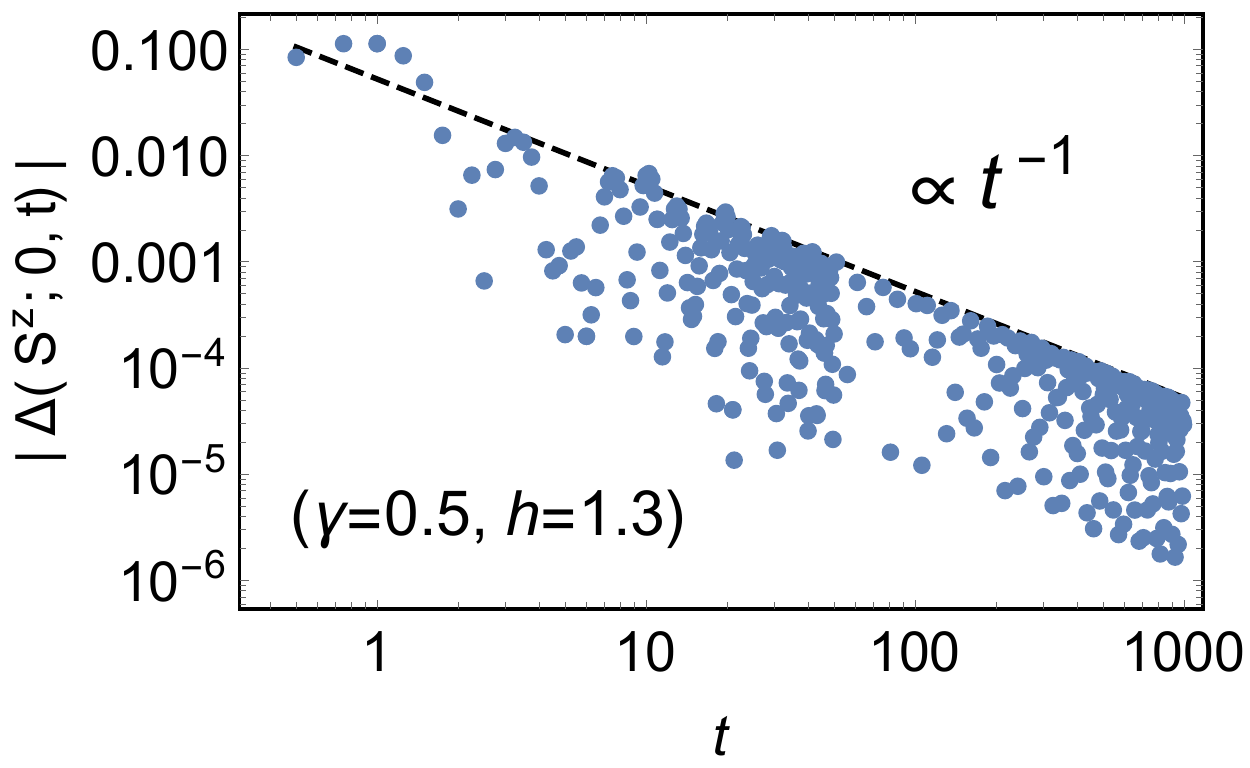}
\label{fig:originDecayC}
}
\subfigure[]{
\centering
\includegraphics[width=0.31\textwidth]{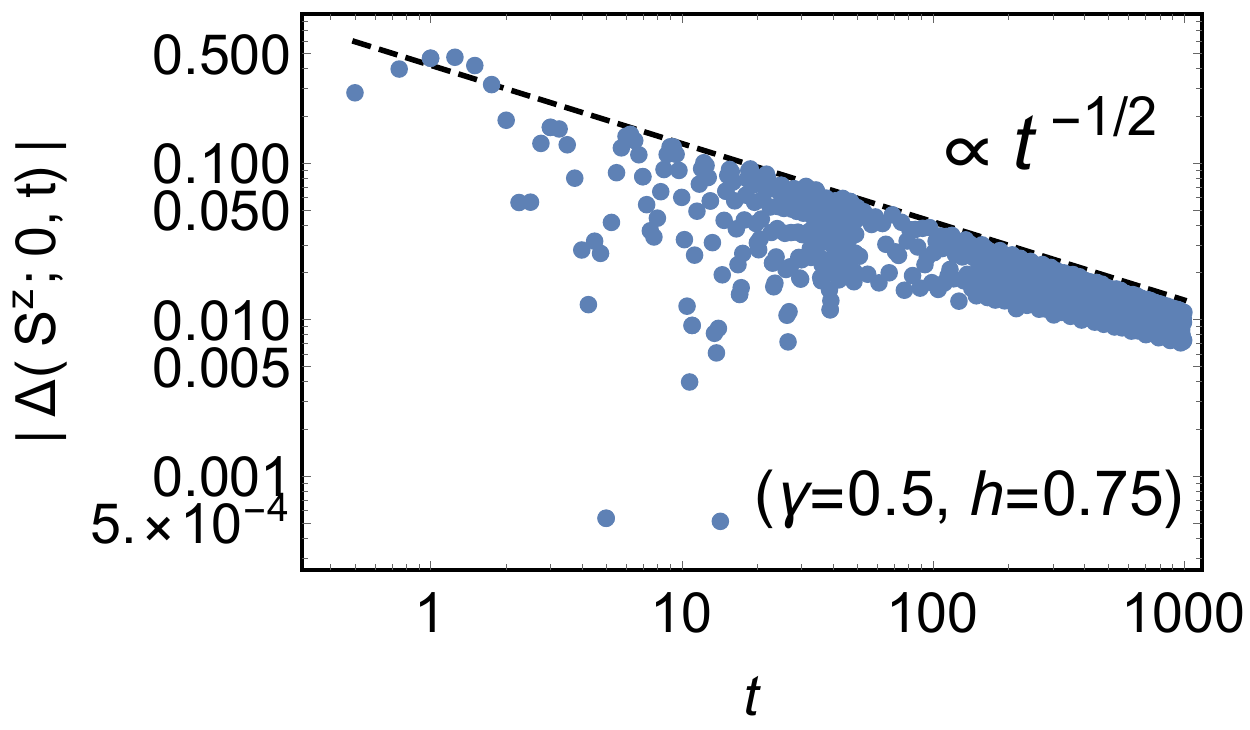}
\label{fig:originDecayB}
}
\caption{Power-law decay of the wave front and the decay at the impacted site $n=0$.
The value of the model parameters, {\it i.e.}, the anisotropy $\gamma$ and the magnetic field $h$ are shown in each panel.
(a), (b) and (c) The blue points show the height of the wave front which forms a light cone at different times, {\it i.e.}, $\Delta(S^{z}; n_{t}^{\rm max}, t)$ for various sets of $t$ and $n_{t}^{\rm max}$, the latter of which denotes the position of the peak of the wave front at time $t$.
(d) and (e) The amplitude of the magnetization change $|\Delta(S^{z}; 0, t)|$ at $n=0$ for different times are plotted as the blue points.
In these cases, the magnetization change decays with oscillations.
In particular, (e) shows the decay of the frozen wave front at $n=0$ in Fig.~\ref{fig:3DB}.
The broken lines in (a)-(e) show the functions which decay as a power law in time $t$, where the exponent is shown in each panel.
In (b), we show the dotted line which decays as $t^{-2/3}$ for comparison in addition to the broken line which decays as $t^{-3/5}$.
The value of the model parameters, {\it i.e.}, the anisotropy $\gamma$ and the magnetic field $h$ are shown in each panel.
}
\label{fig:DecayWave}
\end{figure}

We now focus on the long-time behavior of the wave fronts.
As we have explained in Sec.~\ref{subsec:asympAnaly}, the integral \eqref{eq:Ifunc} decays as a power law in time $t$ as $t^{-1/\kappa}$ in the space-time scaling limit with the integer $\kappa$ determined by Eq.~\eqref{eq:kappa}.
We can estimate the decay of the wave front of the magnetization change $|\Delta(S^{z}; n, t)|$ as $t^{-2/\kappa}$ since it has a quadratic form of the integrals of the form \eqref{eq:Ifunc}.
Figure~\ref{fig:DecayWave} shows the time dependence of the amplitude of a wave front of the magnetization change for five model-parameter sets.
They indeed show power-law decay with various exponents.
Below and in the next section we discuss the origin of these exponents by using the stationary phase analysis.
The decay in Fig.~\ref{fig:waveDecayC} is given in Sec.~\ref{subsec:T23Approx}, the decays in Figs.~\ref{fig:waveDecayH1Gcri} and \ref{fig:waveDecayH1G1} are given in Sec.~\ref{subsec:AiryFail}, and the ones in Fig.~\ref{fig:originDecayC} and \ref{fig:originDecayB} in Sec.~\ref{sec:analyRidge}.

\subsubsection{The decay $t^{-2/3}$ on the wave front}\label{subsec:T23Approx}
Focusing on the parameter region $0<\gamma\leq1$, $0\leq h\leq2$, we find that $\kappa$ is three except for the case of $|h|=|1-\gamma^{2}|$, in which $\kappa$ becomes four with $p^{*}=\pi$, and for the case of $\gamma=\gamma_{c}$, $h=1$, in which $\kappa$ becomes five with $p^{*}=\pi$ (see Sec.~\ref{subsec:AiryFail}).
We thereby find the decay $t^{-2/3}$ in general cases as exemplified in Fig.~\ref{fig:waveDecayC}.
This decay is typically observed in the propagation dynamics.
In Sec.~\ref{sec:Airyin4}, we show that the profile of the magnetization change  around the wave front is well described by using the Airy function.

\subsubsection{The decay $t^{-1/2}$ at the origin}\label{sec:analyRidge}
When the model-parameter set is located on the phase boundary $B$ with $|h|=|1-\gamma^{2}|$, there emerges a frozen wave front (see Fig.~\ref{fig:3DB}), which decays as $t^{-1/2}$, in addition to the propagating wave front, whose decay is well described by $t^{-2/3}$ in the general case of $\kappa=3$.
In this case, the dispersion relation $\varepsilon(p)$ has an inflection point at $p^{**}=\pi$ (see local extrema of $\varepsilon'(p)$ in Fig.~\ref{fig:DispersionVgG05} with $h=0.75$), at which the group velocity $V_{2}=v_{\rm g}(p^{**})=\varepsilon'(p^{**})$ as well as the third derivative of the dispersion vanish, while the forth derivative of the dispersion is given by $d^{4}\varepsilon(p=\pi)/dp^{4}=3/\gamma^2-3$.
Therefore, $\varepsilon^{(4)}(p=\pi)$ is finite and $\kappa$ for this inflection point $p^{**}=\pi$ is four except for the case of $\gamma=1$ and $|h|=|1-\gamma^{2}|=0$, at which the dispersion becomes constant, {\it i.e.}, $\varepsilon(p)=1$.

The result in Fig.~\ref{fig:originDecayB} demonstrates that the amplitude of the frozen wave front with $V_{2}=0$ decays as $t^{-1/2}$ with an oscillation, owing to the decay $t^{-1/4}$ of $\kappa=4$ at $n=0$ of the functions $F,G,Q$, and $W$.
Since the decay $t^{-1/2}$ is slower than $t^{-2/3}$ of the first wave front, the frozen wave front stands out as in Fig.~\ref{fig:3DB}.
In the other parameter regions in the phase diagram, the magnetization change decays as $t^{-1}$ at $n=0$ as is exemplified in Fig.~\ref{fig:originDecayC} (at which the model is located in the region $A$), owing to the decay $t^{-1/2}$ of $\kappa=2$ with an oscillation of the same functions.

The phase boundary $B$ has been identified in some other studies from the viewpoint of the dynamical behavior of the $XY$ model, including the work on a non-equilibrium steady state \cite{NESSxy2008Prosen}, on the relaxation of the magnetization after a global quench \cite{statistical1970XY}, and from a domain wall initial state \cite{Eisler2018hydro}.
Our findings for the frozen wave front suggest that this transition line can be captured by simply observing the frozen wave front around the impacted site after applying a local unitary operation to the system.

\section{The decay $t^{-1}$ and $t^{-3/5}$ in special cases}\label{sec:AiryNew}
In this section, we extend the asymptotic analysis in Sec.~\ref{subsec:asympAnaly}, and analytically discuss the origin of the decay exponents in Figs.~\ref{fig:waveDecayH1Gcri} and \ref{fig:waveDecayH1G1}, as well as of the profiles in Figs.~\ref{fig:ProFileG03H1}--\ref{fig:ProFileG1H1}.

We can approximate the profile of the wave fronts for large $t$ by extending the asymptotic analysis in Eq.~\eqref{eq:Ifuncstation}.
Around the wave front {\it i.e.}, $n\sim vt$ with $v=V_{1}$ and $V_{2}$, the integral \eqref{eq:Ifunc} can be approximated by 
\begin{eqnarray}
I(x,t) = g(p^{*})A_{\kappa}(p^{*},x,t) \mathrm{e}^{it\varepsilon(p^{*}) - ip^{*}x}+ O\left(t^{-2/\kappa}\right)\label{eq:Ifuncapprox}
\end{eqnarray}
as long as
\begin{eqnarray}
{\left|{ \varepsilon^{'}(p^{*})t-x }\right|} \ll \left[\frac{\left|\varepsilon^{(\kappa)}(p^{*})\right|t}{(\kappa-1)!}\right]^{1/\kappa}\label{eq:ValidityAiry}
\end{eqnarray}
(see \ref{sec:APPa} for the derivation), where we define
\begin{eqnarray}
A_{n}(p^{*},x,t) \equiv  \frac{ B_{n}\left(  X_{n} \right) }{\left(\left|\varepsilon^{(n)}(p^{*})\right|t/(n-1)!\right)^{1/n}},\label{eq:funcA}
\end{eqnarray}
\begin{eqnarray}
B_{n}(X)\equiv \int_{-\infty}^{\infty}\frac{dp}{2\pi}\exp\left(ipX+ip^{n}/n\right),
\end{eqnarray}
\begin{eqnarray}
X_{n}\equiv \frac{\varepsilon^{'}(p^{*})t-x}{\left(\left|\varepsilon^{(n)}(p^{*})\right|t/(n-1)!\right)^{1/n}}
\label{eq:DefXn}
\end{eqnarray}
when $\varepsilon^{(n)}(p^{*})>0$.
When $\varepsilon^{(n)}(p^{*})<0$, on the other hand, we change $B_{n}\left(  X_{n} \right)$ in Eq.~\eqref{eq:funcA} to $B^{*}_{n}\left(  -X_{n} \right)$.
(If there are multiple inflection points $p^{*}_{1},\,p^{*}_{2},\,...$ that satisfy $\varepsilon'(p^{*}_{1})=\varepsilon'(p^{*}_{2})=...$, we add up all the contributions from these points, {\it i.e.}, $A_{n}(p^{*}_{1},x,t)+A_{n}(p^{*}_{1},x,t)+...$ .)

Using the approximation \eqref{eq:Ifuncapprox} for Eqs.~\eqref{eq:Fdefint}--\eqref{eq:Wdefint}, we obtain 
\begin{eqnarray}
K_{1}&\simeq (|t_{p^{*}}|^{2}-|s_{p^{*}}|^{2})|t_{p^{*}}|^{2} A_{\kappa}^{2}(p^{*},x,t)
\nonumber\\
&= - \frac{(\cos{p^{*}}+h)}{\varepsilon(p^{*})}\frac{\varepsilon(p^{*}) - (\cos{p^{*}}+h)}{2\varepsilon(p^{*})}A_{\kappa}^{2}(p^{*},x,t),\label{eq:k1app}\\
K_{2}&\simeq (|t_{p^{*}}|^{2}-|s_{p^{*}}|^{2})^2 A_{\kappa}^{2}(p^{*},x,t)
=\frac{(\cos{p^{*}}+h)^{2}}{\left(\varepsilon(p^{*})\right)^{2}}A_{\kappa}^{2}(p^{*},x,t)\label{eq:k2app}
\end{eqnarray}
as the leading behavior of Eq.~\eqref{eq:K12} for large $t$ with $v=V_{1}$ and $V_{2}$, while the next-order term in this approximation is estimated at $O\left(t^{-1/\kappa}\right)\times O\left(t^{-2/\kappa}\right)=O\left(t^{-3/\kappa}\right)$ as a crossing term from the first and second terms in Eq.~\eqref{eq:Ifuncapprox}.
The approximations \eqref{eq:k1app} and \eqref{eq:k2app} are useful as long as $\cos{p^{*}}+h\neq0$.
(See Sec.~\ref{subsec:AiryFail} for the case of $\cos{p^{*}}+h=0$.)

\subsection{Magnetization profile with the Airy function}\label{sec:Airyin4}
\begin{figure}
\begin{indented}\item[]
\includegraphics[width=0.56\textwidth]{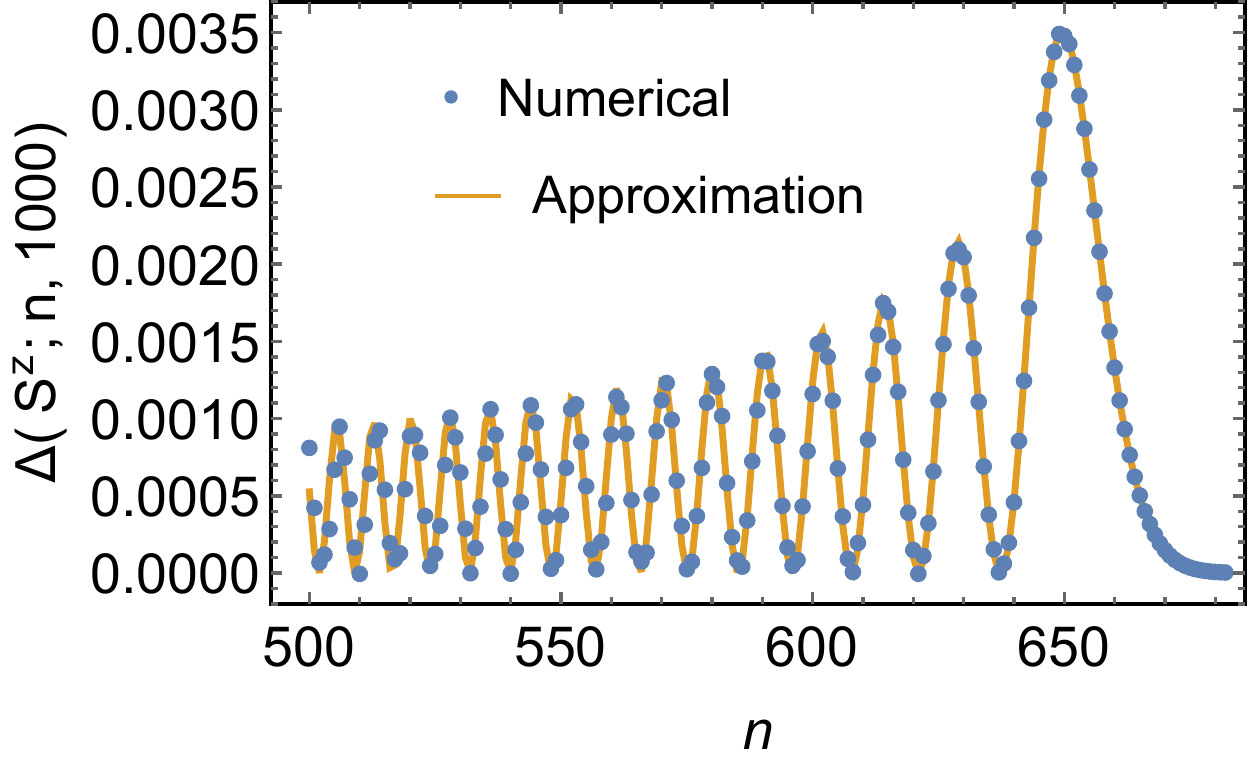}
\caption{The profile of the magnetization change at time $t=1000$ around the light-cone edge for the model parameter $\gamma=0.5$, $h=0.3$, for which the model is located in the region $C$ of Fig.~\ref{fig:phasedig}.
The blue points represent numerical result, while the orange line represents the approximation using Eq.~\eqref{eq:Ifuncapprox} with $\kappa=3$.
The magnetization change $\Delta(S^{z}; n, t)$ has two wave fronts for $n>0$ owing to two inflection points $p^{*}$ and $p^{**}$ of $\varepsilon(p)$ with $v_{g}(p^{*})>v_{g}(p^{**})>0$.
In the approximation, we used Eqs.~\eqref{eq:k1app} and \eqref{eq:k2app} for the first inflection point $p^{*}$ with $\kappa=3$ to calculate the approximated value of $\Delta(S^{z}; n, t)$.
}
 \label{fig:airy}
\end{indented}
\end{figure}

The expressions \eqref{eq:Ifuncapprox} and \eqref{eq:funcA} show not only that the integral $I(x,t)$ decays as $t^{-1/\kappa}$ for large $t$ with $n\sim vt$ as we derived in Sec.~\ref{subsec:asympAnaly}, but also that they well reproduce the profile of the magnetization change of the wave front.
The integral $B_{n}(X)$ can be seen as a generalization of the Airy function of the first kind since $B_{3}(X)={\rm Ai}(X)$.
Figure~\ref{fig:airy} demonstrates a good agreement between the numerical calculation of the magnetization-change profile and the approximation obtained from Eqs.~\eqref{eq:k1app} and \eqref{eq:k2app} with $\kappa=3$.
Although the validity of the approximation is guaranteed only for around $|658-n| \ll 9 $ from Eq.~\eqref{eq:ValidityAiry} in this case, 
the approximation succeeds in describing the profile for a wider region of $n$ in the figure.

This kind of analysis has been performed in several studies, for instance, for the $XX$ model ($\gamma=0$), the Ising model ($\gamma=1$) and the Bose-Hubbard model.
In Refs.~\cite{PRA2012BoseGasPropVelo,Bertini2016edge,XY2018quench}, the wave front of correlation functions after global quenches are argued to be well described in terms of the Airy function ${\rm Ai}(X)=B_{3}(X)$.
In Refs.~\cite{Hunyadi2004XX,Viti2016JointInhomo}, the Airy function is also used to characterize the wave fronts after quenches from step-like inhomogeneous initial states.
On the other hand, to the best of our knowledge, the asymptotic behavior of wave fronts has not been carefully investigated for the $XY$ model with $0<\gamma<1$ so far.

\subsection{The decay $t^{-3/5}$ and $t^{-1}$ in special cases}\label{subsec:AiryFail}

So far we have discussed the cases in which the long-time dynamics of the wave fronts can be well described by the approximation \eqref{eq:Ifuncapprox}.
However, the coefficient $\cos{p^{*}}+h$ in Eqs.~\eqref{eq:k1app} and \eqref{eq:k2app} vanishes when the model-parameter set is located on the line $E$ and on the Ising transition line $h=1$ in the phase diagram and hence the approximation \eqref{eq:Ifuncapprox} is invalidated.
In these parameter regions, the wave fronts show anomalous behavior.

\subsubsection{Anomalous behavior at $h=1$}
For $h=1$, the dispersion relation $\varepsilon(p)$ has an inflection point at $p=\pi$.
Since the right-hand sides of Eqs.~\eqref{eq:k1app} and \eqref{eq:k2app} vanish as in $(\cos{\pi}+1)/\varepsilon(\pi)=\sqrt{(\cos{\pi}+1)/2\gamma^{2}}=0$, the long-time behavior of the wave front owing to this inflection point is given by a higher-order term in the approximation.

For $\gamma>\gamma_{c}=\sqrt{3}/2$ with $h=1$,
we numerically found that the light cone due to this wave front exhibits a peak as we observe in Fig.~\ref{fig:ProFileG1H1}.
The peak height decays as $t^{-1}$ for $h=\gamma=1$ as we show in Fig.~\ref{fig:waveDecayH1G1}.
The decay is consistent with our estimation on the time dependence of the second-order term in the approximations \eqref{eq:k1app} and \eqref{eq:k2app}, namely $O\left(t^{-3/\kappa}\right)$ with $\kappa=3$.

On the other hand, for $0<\gamma<\gamma_{c}$ with $h=1$, 
the dispersion $\varepsilon(p)$ has another inflection point in $0<p<\pi$; see the local extrema of $\varepsilon'(p)$ in Fig.~\ref{fig:DispersionVgH1}.
The wave front corresponding to this new inflection point propagates faster than that of $p^{**}=\pi$ and decays as $t^{-2/3}$, forming a light cone with a clear peak (see Figs.~\ref{fig:ProFileG03H1} and \ref{fig:ProFileG05H1}), whereas the second wave front inside this light-cone region due to the inflection point $p^{**}=\pi$ is expected to decay as $t^{-1}$ as it is the case for $\gamma=1>\gamma_{c}$.
Note that this decay $\Delta(S^{z}; n=vt, t)\sim t^{-1}$ in the space-time scaling limit typically holds inside the light-cone region since the integral $I(vt,t)$ in Eq.~\eqref{eq:Ifunc} behaves as $I(vt,t)\sim t^{-1/2}$ when $|v|$ satisfies $|v|<V_{1}$ and $|v|\neq V_{2}$ so that $\kappa$ takes two.
The second wave front which presumably decays as $t^{-1}$ is hard to identify because it is blurred by the tail of the fastest wave front in this region.

At the point $h=1$ and $\gamma=\gamma_{c}$, namely at the upper edge of $D$ in Fig.~\ref{fig:phasedig}, the other inflection point $p^{*}$ collapses with the inflection point at $p^{**}=\pi$.
At this point, the third and fourth derivatives of the dispersion vanish at $p^{*}=p^{**}=\pi$, while the fifth derivative $\varepsilon^{(5)}(p=\pi)$ survives.
Therefore, despite the leading behavior of the integral Eq.~\eqref{eq:Ifuncapprox} being expected to decay as $\sim t^{-1/5}$, the wave front of the magnetization change shows the decay $\sim t^{-3/5}$ as we have observed in Fig.~\ref{fig:waveDecayH1Gcri}.
Again this decay is consistent with our estimation of the next order of \eqref{eq:k1app} and \eqref{eq:k2app}, $O\left(t^{-3/\kappa}\right)$ with $\kappa=5$.

\subsubsection{Anomalous behavior at $\gamma=1$}\label{sec:AnomalousG1}
In the case of $\gamma=1$, $0<h<1$, namely when the $XY$ model reduces to the transverse Ising model, the coefficients in Eqs.~\eqref{eq:k1app} and \eqref{eq:k2app} again vanish.
Here, we observe that there is no clear peak around the wave front as is exemplified in Fig.~\ref{fig:ProFileG1H07}, whereas there appears a clear peak for $h\geq1$ as is exemplified in Figs.~\ref{fig:ProFileG1H1}--\ref{fig:ProFileG1H15}, and in Fig~\ref{fig:waveDecayH1G1}.
The behavior for $\gamma=1$, $0<h<1$ as well as for $h=1$ is considered to be described by higher-order terms in the approximation \eqref{eq:k1app} and \eqref{eq:k2app}, whose exact form we have not succeeded in obtaining analytically.

\section{Discussion}\label{sec:disc}
In this paper we have investigated the propagation dynamics in the one-dimensional $XY$ model under a magnetic field.
We introduced the local-impact protocol, which is described by a local and instantaneous unitary operation $U_{\rm loc}$ applied to a steady state, 
and focused on the velocity of the propagation and the asymptotic behavior of the amplitude of the propagating wave front.
We found distinctive features of the profile of the magnetization in the $XY$ model, which mediates two prototypical integrable models, the $XX$ chain ($\gamma=0$) and the transverse field Ising chain ($\gamma=1$), particularly in the anti-ferromagnetic phase $0\leq h<1$ as well as in the critical phase $h=1$.

Using numerical calculation and analytical computation, we demonstrated that the model exhibits a frozen wave front and a second wave front inside the light-cone region for $|h|\leq|1-\gamma^{2}|$, namely in the regions $B$ and $C$ in the phase diagram, respectively; see Figs.~\ref{fig:3DB} and \ref{fig:3DA}.
This second wave front only emerges for the anisotropic $XY$ model since it originates from multiple local extrema of the group velocity of the quasiparticles, which can appear only for $\gamma\neq0$ (more specifically, $|h|\leq1-\gamma^{2}$).

We also found that the profile of the magnetization change exhibits drastic difference, that is, the absence of a peak around the wave front (see Fig.~\ref{fig:ProFileG1H07}) for $\gamma=1$, namely on the line $E$ in Fig.~\ref{fig:phasedig}.
While we have provided an analytical description for the origin of this behavior in Sec.~\ref{subsec:AiryFail}, it will be interesting to find a physically relevant explanation using a quasiparticle picture, as well as investigating the universality of this difference in terms of other observables.

The transition line $|h|=|1-\gamma^{2}|$ has been identified in some other studies \cite{NESSxy2008Prosen,statistical1970XY,Eisler2018hydro} from the viewpoint of the dynamical behavior of the $XY$ model.
In our protocol, on the other hand, we can capture this transition line by simply observing the dynamics of the frozen wave front around the impacted site after applying a local unitary operation to the system.
Our results suggest that observing propagation dynamics of the local disturbance in terms of a local spin magnetization can solely show rich and nontrivial behavior of dynamical properties of quantum systems.

For the asymptotic behavior of the propagation dynamics, we have found out that the height of the wave front decays in a power law in time with various exponents depending on the model parameters.
Several other studies have investigated long-time behavior of correlation functions around the light-cone edge under quench protocols.
The Airy function associated with the scaling $t^{-1/3}$ has been used to describe the dynamics around the wave front in order to discuss the asymptote of its height, width and velocity \cite{XY2018quench,Hunyadi2004XX, Bertini2016edge,Viti2016JointInhomo,PRA2012BoseGasPropVelo},
and the scaling $t^{-1/3}$ and its square $t^{-2/3}$ appeared universally in light-cone dynamics.
In the present paper, in contrast, we have revealed using the local-impact protocol that the scaling for the height of the wave front around the light-cone edge can be given not only by $t^{-2/3}$ but also by $t^{-1}$ and $t^{-3/5}$ depending on the parameter values (see Fig.~\ref{fig:DecayWave}), by carefully investigating the dispersion relation $\varepsilon(p)$ and the coefficient for the approximation.
In particular, we found that the leading terms \eqref{eq:k1app} and \eqref{eq:k2app} in the approximation of $\Delta(S^{z}; n, t)$ vanish when the model is on the line $h=1$ or on the line $0<h<1$ and $\gamma=1$, for which the relation $\cos{p^{*}}+h=0$ holds for a local extremum $p=p^{*}$ of $v_{\rm g}(p)$.

The local-impact protocol which we introduced in this paper may provide a new insight into the study of dynamics in isolated quantum systems.
It will be important to investigate the propagation dynamics in this protocol in terms of other observables, such as the magnetization in the $x$ directions and the entanglement entropy as have been studied in Refs.~\cite{divakaran2011non,zauner2015time} for the transverse Ising model.
Studying a relaxation process after applying the local impacts for all sites is an interesting direction for future research.

Recently, we became aware of an independent work Ref.~\cite{PRR2019caustics}, which has considered a similar setting, namely creation of a single quasiparticle at the origin of the anisotropic $XY$ model, in order to discuss the similarity between light-cone behavior in spin chains and quantum caustics.
It mainly considered a localized quasiparticle excitation, {\it i.e.}, $(1/\sqrt{L})\sum_{k}\eta^{\dagger}_k |GS\rangle$, which is a rather nongeneric initial condition, and found the existence of edges of the second light cone and collapse of them at $|h|=|1-\gamma^{2}|$, which is consistent to our results, whereas the absence of the peak on the line $E$ has not been observed.
We consider the singular behaviors at $h=1$, $\gamma=\gamma_c$ as well as on the line $|h|=|1-\gamma^{2}|$ and $h=1$ that we found in the present paper to be universal because they arise from the singular properties of the dispersion relation, whereas the robustness of the behavior on the line $E$ against the initial state remains an interesting question.

After completion of the present manuscript, we were notified that Ref.~\cite{Kormos2017Referee2FIVE} considered the wave-front dynamics in the case of the transverse field Ising model ($\gamma=1$) and the $XX$ model ($\gamma=0$) under the domain-wall initial condition, using a similar asymptotic analysis.
The point that is made there but is missing in the present paper is the absence of even the second-order term in the approximation for $\gamma=1$; we did not check it in \ref{sec:AnomalousG1}.
We stress here that this does not occur for $\gamma<1$.
\ack
The author is grateful to Naomichi Hatano for fruitful discussions and carefully proofreading the manuscript.
The author also thanks Akira Shimizu, Synge Todo, Yuichiro Matsuzaki, Yuya Seki, and Lee Jaeha for valuable comments. 

\appendix
$\,$
\section{Derivation of Eqs.~\eqref{eq:Fdefint}--\eqref{eq:Qdefintxx} and irrelevance of the degeneracy}\label{sec:AppA}
We here explain the derivation of Eqs.~\eqref{eq:Fdefint}--\eqref{eq:Qdefintxx} and give details of our statement in Sec.~\ref{subsec:digonal} that the choice of our ground states of block-diagonalized Hamiltonians is irrelevant to the evaluation of Eq.~\eqref{eq:DeltaSzForm}.
We rewrite the magnetization change \eqref{eq:DeltaSzForm} as
\begin{eqnarray}
\Delta(S^{z}; n, t) &=\langle{U_{\rm loc}^{\dagger} S_{n}^{z} (t)U_{\rm loc} }\rangle_{} - \langle{S_{n}^{z}(0)  }\rangle_{}\nonumber \\
&=\langle U_{\rm loc}^{\dagger} [S_{n}^{z} (t), U_{\rm loc}]   \rangle_{},\label{eq:appDerivDelta1}
\end{eqnarray}
using $\langle S_{n}^{z} (t)\rangle_{}=\langle S_{n}^{z} (0)\rangle_{}$.
Since $[P_{\pm}, S_{n}^{z}]=[P_{\pm}, U_{\rm loc}]=0$, the operator in the right-hand side of Eq.~\eqref{eq:appDerivDelta1} acts independently on the two sectors defined by $P_{\pm}$:  
\begin{eqnarray}
U_{\rm loc}^{\dagger} [S_{n}^{z} (t), U_{\rm loc}] = \sum_{a=\pm}P_{a}U_{\rm loc}^{\dagger} [\mathrm{e}^{iH^{a}t} S_{n}^{z} \mathrm{e}^{-iH^{a}t}, U_{\rm loc}] P_{a}. \label{eq:AppPdeltaP}
\end{eqnarray}
Therefore, we can parallelly calculate $\Delta(S^{z}; n, t)$ for the ground state in the two sectors.
We note that the ground states of the $XY$ model can be $\ket{\rm GS}_{+}$, $\ket{\rm GS}_{-}$, or a superposition of them depending on the size $L$, the anisotropy $\gamma$, and the field $h$; see Ref.~\cite{okuyama2015anomalous}.

We first show that the anti-commutators $\{c_n^{\dagger}(t), c_m\}$ and $\{c_n(t), c_m\}$ are c-numbers, and then derive the expressions of $F$, $G$, $Q$, and $W$ in Eqs.~\eqref{eq:Fdefint}--\eqref{eq:Qdefintxx}.
From the equation of motion of quasiparticles $\eta_{p}$,  {\it i.e.}, $(d/dt)\eta_{p}(t)=i[H^{a},\eta_{p}(t)]=-i\varepsilon(p)\eta_{p}(t)$ with respect to each sector $a=\pm$, we obtain $\eta_{p}(t)=\eta_{p}\exp\left[(-i\varepsilon(p))t\right]$, and thereby obtain the expression of the Jordan-Wigner fermions $c_{n}(t)$ in terms of $\eta_{p}$ as
\begin{eqnarray}
c_n(t)=\mathrm{e}^{iH^{a}t}c_{n}\mathrm{e}^{-iH^{a}t}
=\frac{1}{\sqrt{L}} {\sum_{p}}^{a}(s_{p} \mathrm{e}^{-i\varepsilon(p) t} \eta_{p} + t_{p} \mathrm{e}^{i\varepsilon(p) t} \eta_{-p}^{\dagger})\mathrm{e}^{-ipn}\label{eq:appCnT}
\end{eqnarray}
from Eq.~\eqref{eq:etapsptp}.
Using Eq.~\eqref{eq:appCnT} and the relations $s_{p}=s_{-p}=s_{p}^{*}$, $\varepsilon(p)=\varepsilon(-p)$ and $t_{p}=-t_{-p}=-t_{p}^{*}$, we obtain
\begin{eqnarray}
\fl
\{c_n^{\dagger}(t), c_m\} =\left\{ \frac1{\sqrt{L}}{\sum_{p}}^{a} (s_{p}\mathrm{e}^{i\varepsilon(p)t} {\eta^{\dagger}_p} -t_{p}\mathrm{e}^{-i\varepsilon(p)t} {\eta_{-p}})\mathrm{e}^{ipn} ,\, \frac1{\sqrt{L}}{\sum_{q}}^{a} (s_{q} {\eta_q} + t_{q} \eta_{-q}^{\dagger})\mathrm{e}^{-iqm} \right\}\nonumber\\
= \frac1L {\sum_{p,q}}^{a}\left(  s_{p}s_{q}\mathrm{e}^{i\varepsilon(p)t} \{ \eta_{p}^{\dagger},\,\eta_{q}\} - t_{p}t_{q}\mathrm{e}^{-i\varepsilon(p)t} \{ \eta_{-p},\,\eta_{-q}^{\dagger} \}\right.\nonumber\\
\,\,\,\,\,\,\left.+s_{p}t_{q}\mathrm{e}^{i\varepsilon(p)t} \{ \eta_{p}^{\dagger},\,\eta_{-q}^{\dagger}\} - t_{p}s_{q}\mathrm{e}^{-i\varepsilon(p)t} \{ \eta_{-p},\,\eta_{q}\} \right)\mathrm{e}^{i(pn-qm)}\nonumber\\
= \frac1L {\sum_{p}}^{a}\left(  s_{p}^{2}\mathrm{e}^{i\varepsilon(p)t}  - t_{p}^{2} \mathrm{e}^{-i\varepsilon(p)t} \right)\mathrm{e}^{ip(n-m)}\nonumber\\
= \frac1L {\sum_{p}}^{a}\left(  |s_{p}|^{2}\Phi_{p}(n-m,t)+ |t_{p}|^{2} \Phi^{*}_{p}(n-m,t)\right),\label{eq:appFxy1}
\end{eqnarray}
\begin{eqnarray}
\fl
\{c_n(t), c_m\} =\left\{ \frac{1}{\sqrt{L}} {\sum_{p}}^{a}(s_{p} \mathrm{e}^{-i\varepsilon(p) t} \eta_{p} + t_{p} \mathrm{e}^{i\varepsilon(p) t} \eta_{-p}^{\dagger})\mathrm{e}^{-ipn} ,\,  \frac{1}{\sqrt{L}}  {\sum_{q}}^{a}(s_{q}\eta_{q} + t_{q}  \eta_{-q}^{\dagger})\mathrm{e}^{-iqm} \right\}\nonumber\\
= \frac1L {\sum_{p,q}}^{a}\left(s_{p}t_{q}\mathrm{e}^{-i\varepsilon(p)t} \{ \eta_{p},\,\eta_{-q}^{\dagger}\}  + t_{p}s_{q}\mathrm{e}^{i\varepsilon(p)t} \{ \eta_{-p}^{\dagger},\,\eta_{q}\}  \right)\mathrm{e}^{-i(pn+qm)}\nonumber\\
= \frac1L {\sum_{p}}^{a}\left(-s_{p}t_{p}\mathrm{e}^{-i\varepsilon(p)t}  + t_{p}s_{p}\mathrm{e}^{i\varepsilon(p)t}   \right)\mathrm{e}^{-ip(n-m)}\nonumber\\
= \frac1L {\sum_{p}}^{a} s_{p}t_{p} \left(\Phi_{p}(n-m,t)+\Phi^{*}_{p}(n-m,t)\right),\label{eq:appGxy1}
\end{eqnarray}
where
\begin{eqnarray}
\Phi_{p}(n,t)= \mathrm{e}^{i(\varepsilon(p)t-pn)},
\end{eqnarray}
for the anisotropic case $\gamma\neq0$, and
\begin{eqnarray}
\{c_n^{\dagger}(t), c_m\} &=\frac1{\sqrt{L}}{\sum_{p,q}}^{a}\mathrm{e}^{i(\cos{p}+h)t}\{\eta_{p}^{\dagger} , \eta_{q}\} \mathrm{e}^{ipn-iqm}\nonumber\\
&= \frac1{\sqrt{L}}{\sum_{p}}^{a}\mathrm{e}^{i(\cos{p}+h)t+ip(n-m)},\label{eq:appFxx1}\\
\{c_n(t), c_m\} &=\,0\label{eq:appGxx1}
\end{eqnarray}
for the isotropic case $\gamma=0$.
Equations~\eqref{eq:appFxy1}--\eqref{eq:appGxx1} clearly show that the anti-commutators $\{c_n^{\dagger}(t), c_0\}$ and $\{c_n(t), c_0\}$ are c-numbers, which we denote by $F=F(n,t)$ and $G=G(n,t)$, respectively, as in Eqs.~\eqref{eq:defFGbraket} and \eqref{eq:defFGbraket2}.
In Eqs.~\eqref{eq:appFxx1} and \eqref{eq:appGxx1}, we used an expression $c_n(t)=(\sqrt{1/L}) {\sum_{p}}^{\!\!\!a} \eta_{p} \exp\left[-i\varepsilon(p) t - ipn\right]$ for $\gamma=0$ since $s_{p}\equiv1$ and $t_{p}\equiv0$ from the definitions \eqref{eq:sp} and \eqref{eq:sptp}.
We obtain the expression in \eqref{eq:DeltaSzForm} with \eqref{eq:K12} by utilizing the fact that Eqs.~\eqref{eq:appFxy1} and \eqref{eq:appGxy1} are c-numbers.

Then Eqs.~\eqref{eq:appFxy1} and \eqref{eq:appGxy1} immediately yield $F$ and $G$.
We find $Q$ and $W$ by additionally taking into account Eqs.~\eqref{eq:gsXY} for $\gamma\neq0$, and Eq.~\eqref{eq:gsXX} for $\gamma=0$ in calculating the expectation values with respect to the ground state:
\begin{eqnarray}
\langle  c_n^{\dagger}(t) c_0 \rangle_{a} &= \frac1L {\sum_{p,q}}^{a}\left(  s_{p}s_{q}\mathrm{e}^{i\varepsilon(p)t} \langle  \eta_{p}^{\dagger}\eta_{q}\rangle_{a} - t_{p}t_{q}\mathrm{e}^{-i\varepsilon(p)t}\langle \eta_{-p}\eta_{-q}^{\dagger} \rangle_{a} \right)\mathrm{e}^{ipn}\nonumber\\
&= \frac1L {\sum_{p}}^{a} - t_{p}^{2} \mathrm{e}^{-i\varepsilon(p)t} \mathrm{e}^{ipn}=  \frac1L {\sum_{p}}^{a} |t_p|^{2} \Phi^{*}_{p}(n,t),
\end{eqnarray}
\begin{eqnarray}
\langle c_n(t) c_0 \rangle_{a} &= \frac1L {\sum_{p,q}}^{a}\left( s_{p}t_{q}\mathrm{e}^{-i\varepsilon(p)t} \langle  \eta_{p}\eta_{-q}^{\dagger}\rangle_{a} + t_{p}s_{q}\mathrm{e}^{i\varepsilon(p)t} \langle  \eta_{-p}^{\dagger}\eta_{q} \rangle_{a}  \right)\mathrm{e}^{-ipn}\nonumber\\
&=  \frac1L {\sum_{q}}^{a}  s_{-q}t_{q}\mathrm{e}^{-i\varepsilon(p)t} \mathrm{e}^{inq}= \frac1L {\sum_{p}}^{a} s_{p}t_{p}\Phi^{*}_{p}(n,t)
\end{eqnarray}
for the anisotropic case $\gamma\neq0$, and
\begin{eqnarray}
Q(n,0,t) &=\,\langle  c_n^{\dagger}(t) c_0 \rangle_{a}\,=\frac1L{\sum_{p,q}}^{a}\mathrm{e}^{i(\cos{p}+h)t}\langle  \eta_{p}^{\dagger}  \eta_{q}\rangle_{a}\mathrm{e}^{ipn}\nonumber\\
&= \frac1L{\sum_{{p;  \cos{p}+h\leq0}}}^{ a}\,\mathrm{e}^{i(\cos{p}+h)t+ipn},
\end{eqnarray}
\begin{eqnarray}
W(n,0,t)=\,\langle  c_n(t) c_0 \rangle_{a}\,=\frac1{L}{\sum_{p,q}}^{a}\mathrm{e}^{i(\cos{p}+h)t}\langle  \eta_{p}  \eta_{q}\rangle_{a}\mathrm{e}^{-ipn}=0
\end{eqnarray}
for the isotropic case $\gamma=0$, 
where the angular brackets $\langle\cdots \rangle_{a}$ denote the expectation value with respect to the ground state $\ket{\rm GS}_{a}$.

We arrive at the expressions \eqref{eq:Fdefint}--\eqref{eq:Qdefintxx} by taking the thermodynamic limit $L\rightarrow\infty$ to replace the sum $\sum_{p}^{a}$ over 
$p=2\pi j /L$ with the integral $\int_{-\pi}^{\pi}dp$, where $j=-(L-1)/2, ... ,-1/2, 1/2, ... ,(L-1)/2$ for the Neveu-Schwarz sector $a=+$ and $j=-L/2-1, ... ,-1,0,1, ... ,L/2$ for the Ramond sector $a=-$.

Now we show that the choice of the ground state is irrelevant to the calculation of the magnetization change $\Delta(S^{z}; n, t)$, {\it i.e.}, the difference between
\begin{eqnarray}
{}_{+}\bra{\rm GS}U_{\rm loc}^{\dagger}[S^{z}_{n}(t), U_{\rm loc}]\ket{\rm GS}_{+}\label{eq:appGSplus}
\end{eqnarray}
and
\begin{eqnarray}
{}_{-}\bra{\rm GS}U_{\rm loc}^{\dagger}[S^{z}_{n}(t), U_{\rm loc}] \ket{\rm GS}_{-} \label{eq:appGSminus}
\end{eqnarray}
can be ignored in the thermodynamic limit.
The difference only comes from the way in which we take the sum over $p$ before taking the thermodynamic limit in order to obtain the integral representations in Eqs.~\eqref{eq:Fdefint}--\eqref{eq:Wdefint}, \eqref{eq:Qdefintxx}, and \eqref{eq:Fdefintxx}.
Since the correction for replacing a discrete sum over $p$ for the integral over $[-\pi, \pi]$ is estimated at $O\left(L^{-1}\right)$, the difference is irrelevant in the thermodynamic limit.
Therefore, we do not have to specify which sector the ground state of \eqref{eq:H} belongs to in the calculation of the magnetization change $\Delta(S^{z}; n, t)$.

\section{Comparison with the Lieb-Robinson velocity}\label{sec:AppLR}
\begin{figure}
\begin{indented}\item[]
\centering
\includegraphics[width=0.5\textwidth]{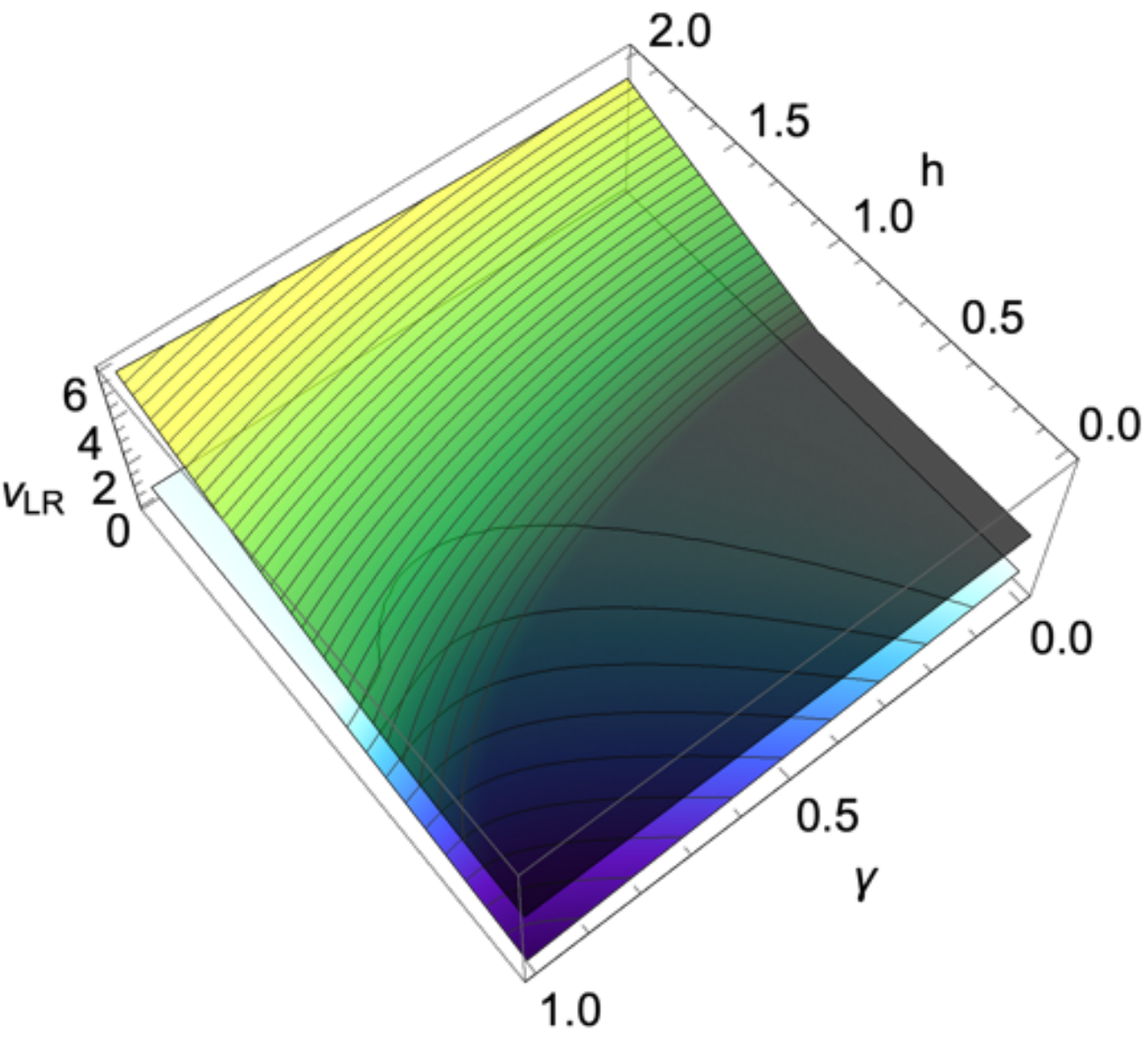}
 \caption{The Lieb-Robinson velocity $v_{\rm LR}$ (the green surface) and the maximum group velocity $V_{1}$ (the blue surface) for the $XY$ model with the anisotropy $\gamma$ and the magnetic field $h$.}
\label{fig:LRcomp}
\end{indented}
\end{figure}
As we mentioned in Introduction, the Lieb-Robinson bound \cite{lieb1972finite} provides a bound for the velocity of the information propagation in lattice spin systems with local interactions.
However, the Lieb-Robinson velocity depends only on the operator norm of the local Hamiltonian, particularly in one-dimensional systems with nearest-neighbor interaction \cite{hastings2010locality}.
The characteristic velocity for the propagation dynamics in a given system may generally depend nontrivially on the property of the system.

The Lieb-Robinson bound in one-dimensional models with nearest-neighbor interaction is expressed as follows:
\begin{eqnarray} 
\| [ A, B(t)] \| \leq2 |X||Y| \| A \| \| B \| \mathrm{e}^{ v_{\rm LR} | t |-d_{(X,Y)} },\label{eq:appLR}\\
v_{\rm LR}:=2\mathrm{e} \max_{n} \left\|  H_{n} \right\|, \label{eq:appLRvelo}
\end{eqnarray}
where $A$ and $B$ are local observables in the region $X$ and $Y$, respectively, and we denote: the unitary time evolution of $B$ over a period $t$ by $B(t)$; the number of sites included in the regions by $|\cdot|$; the distance between $X$ and $Y$ on the chain by $d_{(X,Y)}$; the constant $\exp(1)\sim2.718...$ by $ \mathrm{e}$; the operator norm of a local Hamiltonian of the model $H=\sum_{n}H_{n}$ by $\left\|  H_{n} \right\|$.
The inequality \eqref{eq:appLR} shows that the operator norm of the commutator between observables is exponentially suppressed when $v_{\rm LR} | t | < d_{(X,Y)}$.
We therefore refer to $v_{\rm LR}$ as the Lieb-Robinson velocity.
The local Hamiltonian for the $XY$ model is given by $H_{n}=(1+\gamma)S_n^xS_{n+1}^x +(1-\gamma)S_n^yS_{n+1}^y+(h/2) (S_n^z+S_{n+1}^z)$, and hence the Lieb-Robinson velocity \eqref{eq:appLRvelo} in this model reduces to
\begin{eqnarray}
v_{\rm LR}= 2\mathrm{e}\left\|  H_{n} \right\|  =\max\left[ \mathrm{e},  \mathrm{e}\sqrt{h^{2}+\gamma^{2}}\right].
\end{eqnarray}

In Fig.~\ref{fig:LRcomp}, we compare the Lieb-Robinson velocity $v_{\rm LR}$ and the maximum group velocity $V_{1}$ (see Fig.~\ref{fig:Ep_Max_Velo(527)}) in terms of the dependence on the model parameters.
We can see that $V_{1}$ is much less than $v_{\rm LR}$ and the dependence on the model parameters is also different.

\section{Derivation of Eq.~\eqref{eq:Ifuncapprox}}\label{sec:APPa}
We here present the derivation of Eq.~\eqref{eq:Ifuncapprox} in Sec.~\ref{sec:AiryNew}.
We focus on the integral around the inflection point $p^{*}$ of the dispersion relation $\varepsilon(p)$, which gives the leading contribution of $I(x,t)$ for $t\gg1$ and $\varepsilon'(p^{*})t-x=v_{\rm g}(p^{*})t-x\sim0$.
First we expand $\varepsilon(p)$ around $p^{*}$ as
\begin{eqnarray}
\varepsilon(p)\simeq\varepsilon(p^{*}) + \varepsilon'(p^{*})(p-p^{*}) + \frac1{\kappa!}\varepsilon^{(\kappa)}(p^{*})(p-p^{*})^{\kappa}.
\end{eqnarray}
Then a straightforward calculation yields
\begin{eqnarray}
\fl
I(x,t)= \int_{-\pi}^{\pi}\frac{dp}{2\pi} g(p)\mathrm{e}^{i \varepsilon(p)t -i x p} \label{eq:APPIxtfirstTOP} \\
\fl
\,\,\,\,\,\,\,\,\,\,\,\,\,\,\,\,\,\,
\simeq \int_{-\pi}^{\pi}\frac{dp}{2\pi} g(p^{*})\exp\left[{i \left(\varepsilon(p^{*}) + \varepsilon'(p^{*})(p-p^{*}) + \frac1{\kappa!}\varepsilon^{(\kappa)}(p^{*})(p-p^{*})^{\kappa}  \right)t -i x p}\right]\nonumber\\
\fl
\,\,\,\,\,\,\,\,\,\,\,\,\,\,\,\,\,\,
\simeq g(p^{*})\mathrm{e}^{i \varepsilon(p^{*})t -i x p^{*}} \int_{-\infty}^{\infty}\frac{dp}{2\pi} \exp\left[i \left(\varepsilon'(p^{*})(p-p^{*}) + \frac1{\kappa!}\varepsilon^{(\kappa)}(p^{*})(p-p^{*})^{\kappa} \right)t \right. \nonumber\\
\fl
\,\,\,\,\,\,\,\,\,\,\,\,\,\,\,\,\,\,\,\,\,\,\,\,
\left.-i x (p-p^{*})\right] \label{eq:APPIxtfirstTOP3}\\
\fl
\,\,\,\,\,\,\,\,\,\,\,\,\,\,\,\,\,\,
=g(p^{*})\mathrm{e}^{i \varepsilon(p^{*})t -i x p^{*}} \int_{-\infty}^{\infty}\frac{dq}{2\pi} \exp\left[i \left(\varepsilon'(p^{*})t - x \right) q + it\frac1{\kappa!}\varepsilon^{(\kappa)}(p^{*})q^{\kappa} \right]\nonumber\\
\fl
\,\,\,\,\,\,\,\,\,\,\,\,\,\,\,\,\,\,
=g(p^{*})\mathrm{e}^{i \varepsilon(p^{*})t -i x p^{*}} \left(\frac{(\kappa-1)!}{\left|\varepsilon^{(\kappa)}(p^{*})\right|t}\right)^{1/\kappa}\int_{-\infty}^{\infty}\frac{d{\tilde q}}{2\pi}\exp\left( i X_{\kappa}  {\tilde q}+\frac{i}{\kappa}{\tilde q}^{\kappa}\right)\label{eq:APPIxtfirstBotom2}\\
\fl
\,\,\,\,\,\,\,\,\,\,\,\,\,\,\,\,\,\,
= g(p^{*}) A_{\kappa}(p^{*},x,t) \mathrm{e}^{it\varepsilon(p^{*}) - ip^{*}x}\label{eq:APPIxtfirst}
\end{eqnarray}
with
\begin{eqnarray}
A_{n}(p^{*},x,t) \equiv {\tilde C}_{n} t^{-1/n} B_{n}\left(  X_{n} \right),\label{eq:APPfuncA}\\
B_{n}(X) \equiv \int_{-\infty}^{\infty}\frac{d{\tilde q}}{2\pi}\exp\left[i{\tilde q}X+i{\tilde q}^{n}/n\right],\label{eq:AppDefBn}\\
X_{n} \equiv \frac{\varepsilon^{'}(p^{*})t-x}{\left(\left|\varepsilon^{(n)}(p^{*})\right|t/(n-1)!\right)^{1/n}},\\
{\tilde C}_{n} \equiv \left( \frac{(n-1)!}{ | \varepsilon^{(n)}(p^{*}) | } \right)^{1/n}
\end{eqnarray}
as in Eqs.~\eqref{eq:Ifuncapprox}--\eqref{eq:DefXn}.
Here we assumed $\varepsilon^{(n)}(p^{*})>0$ and $p^{*}\in (-\pi,\pi)$.
For $\varepsilon^{(n)}(p^{*})<0$, we change $B_{n}\left(  X_{n} \right)$ in Eq.~\eqref{eq:AppDefBn} to $B^{*}_{n}\left(  -X_{n} \right)$.
In the line \eqref{eq:APPIxtfirstTOP3} we extended the integration region since the contribution from the integral region far from $p=p^{*}$ is small for $t\gg1$ with $\varepsilon'(p^{*})t-x\sim0$.
If $p^{*}$ is on one of the boundaries of the integration region ({\it i.e.}, $p^{*}=\pm \pi$), which is the case for the frozen wave front on the line $B$ and for the second wave front on the line $D$, we extend the integration region as $\pm\int_{\mp\infty}^{0}dp$ instead of $\int_{-\infty}^{\infty}dp$ in the line \eqref{eq:APPIxtfirstTOP3} and thereafter.
In the line \eqref{eq:APPIxtfirstBotom2} we changed the variable of  integration $q=p-p^{*}$ with
\begin{eqnarray}
{\tilde q} = \left( t\frac{ | \varepsilon^{(\kappa)}(p^{*}) | }{(\kappa-1)!} \right)^{1/\kappa} q.
\end{eqnarray}
The final result \eqref{eq:APPIxtfirst} of this approximation works well for large $t$ and $\varepsilon^{'}(p^{*})t-x \ll \left[\left|\varepsilon^{(\kappa)}(p^{*})\right|t/(\kappa-1)!\right]^{1/\kappa}$.

Next, generalizing the treatment in Ref.~\cite{perfetto2017ballistic}, we estimate the second-order term in this approximation.
It is obtained by taking the higher-order terms in the expansion into account.
We expand $g(p)$ and $\varepsilon(p)$ around $p^{*}$ as follows:
\begin{eqnarray}
g(p) \simeq g(p^{*})+ \frac{g^{(\xi)}(p^{*}) }{\xi !}(p-p^{*})^{\xi},\\
\fl 
\varepsilon(p) \simeq\varepsilon(p^{*}) + \varepsilon'(p^{*})(p-p^{*}) + \frac{\varepsilon^{(\kappa)}(p^{*})}{\kappa!}(p-p^{*})^{\kappa} + \frac{\varepsilon^{(\lambda)}(p^{*})}{\lambda!}(p-p^{*})^{\lambda}.
\end{eqnarray}
Substituting these expansion for \eqref{eq:APPIxtfirstTOP} and performing estimation in the same manner as in Eqs.~\eqref{eq:APPIxtfirstTOP}--\eqref{eq:APPIxtfirstBotom2},
we obtain
\begin{eqnarray}
\fl
I(x,t) \simeq\, \mathrm{e}^{i \varepsilon(p^{*})t -i x p^{*}} \int_{-\infty}^{\infty}\frac{dq}{2\pi} (g(p^{*})+ \frac1{\xi !}g^{(\xi)}(p^{*})q^{\xi})\exp\left[i \left(\varepsilon'(p^{*})t - x \right) q + it\frac1{\kappa!}\varepsilon^{(\kappa)}(p^{*})q^{\kappa}  \right]
\nonumber\\
\times\exp\left( it\frac1{\lambda!}\varepsilon^{(\lambda)}(p^{*})q^{\lambda}  \right)\nonumber\\
\fl
\,\,\,\,\,\,\,\,\,\,\,\,\,\,\,\,\,\,=\,  \mathrm{e}^{i \varepsilon(p^{*})t -i x p^{*}} {\tilde C}_{\kappa}t^{-1/\kappa}\int_{-\infty}^{\infty}\frac{d{\tilde q}}{2\pi} \left[g(p^{*})+ \frac{g^{(\xi)}(p^{*})}{\xi!} \left(t^{-1/\kappa} {\tilde C}_{\kappa}{\tilde q}\right)^{\xi}\right]\exp\left(i X_{\kappa} {\tilde q} + i\frac1{\kappa}{\tilde q}^{\kappa}  \right)
\nonumber\\
\fl
\,\,\,\,\,\,\,\,\,\,\,\,\,\,\,\,\,\,\,\,\,\,\,\,
\times\exp\left[  it  \frac{\varepsilon^{(\lambda)}(p^{*})}{\lambda!} \left(t^{-1/\kappa} {\tilde C}_{\kappa}{\tilde q}\right)^{\lambda}  \right]\nonumber\\
\fl
\,\,\,\,\,\,\,\,\,\,\,\,\,\,\,\,\,\,
=\,  t^{-1/\kappa} \mathrm{e}^{i \varepsilon(p^{*})t -i x p^{*}} \int_{-\infty}^{\infty}\frac{d{\tilde q}}{2\pi} \left[ g(p^{*}){\tilde C}_{\kappa} + C_{2,g} {\tilde q}^{\xi} t^{-\xi/\kappa} + C_{2,\varepsilon} {\tilde q}^{\lambda} t^{-(\lambda-\kappa)/\kappa} \right]
\nonumber\\
\fl
\,\,\,\,\,\,\,\,\,\,\,\,\,\,\,\,\,\,\,\,\,\,\,\,
\times\exp\left( i X{\tilde q}  + i\frac1{\kappa}{\tilde q}^{\kappa} \right)\label{eq:APPIxtSeclast}\\
\fl
\,\,\,\,\,\,\,\,\,\,\,\,\,\,\,\,\,\,
=\, g(p^{*})A_{\kappa}(p^{*},x,t)\mathrm{e}^{it\varepsilon(p^{*}) - ip^{*}x}\nonumber\\
\fl
\,\,\,\,\,\,\,\,\,\,\,\,\,\,\,\,\,\,\,\,\,\,\,\,
+
t^{-(1+\xi)/\kappa} C_{2,g}\mathrm{e}^{i \varepsilon(p^{*})t -i x p^{*}} \int_{-\infty}^{\infty}\frac{d{\tilde q}}{2\pi}  {\tilde q}^{\xi}  \exp\left( i X{\tilde q}  + i\frac1{\kappa}{\tilde q}^{\kappa} \right) \nonumber \\
\fl
\,\,\,\,\,\,\,\,\,\,\,\,\,\,\,\,\,\,\,\,\,\,\,\,
+
t^{-(1+\lambda-\kappa)/\kappa} C_{2,\varepsilon}\mathrm{e}^{i \varepsilon(p^{*})t -i x p^{*}} \int_{-\infty}^{\infty}\frac{d{\tilde q}}{2\pi}  {\tilde q}^{\lambda}  \exp\left(i X{\tilde q}  + i\frac1{\kappa}{\tilde q}^{\kappa} \right)\label{eq:APPIxtSeclastLAST}
\end{eqnarray}
up to $O\left(t^{-(1+\xi+\lambda-\kappa)/\kappa}\right)$,
where we defined the constants $C_{2,\varepsilon}$ and $ C_{2,g}$ as
\begin{eqnarray}
C_{2,\varepsilon}&=i\frac{\varepsilon^{(\lambda)}(p^{*})}{\lambda!} {\tilde C}_{\kappa}^{\lambda+1},\\
C_{2,g}&=\frac{g^{(\xi)}(p^{*})}{\xi!} {\tilde C}_{\kappa}^{\xi+1}.
\end{eqnarray}
In the line \eqref{eq:APPIxtSeclast} we used the approximation 
\begin{eqnarray}
\fl
\exp\left[  it  \frac{\varepsilon^{(\lambda)}(p^{*})}{\lambda!} \left(t^{-1/\kappa} {\tilde C}_{\kappa}{\tilde q}\right)^{\lambda}  \right] = 1+ it  \frac{\varepsilon^{(\lambda)}(p^{*})}{\lambda!} \left(t^{-1/\kappa} {\tilde C}_{\kappa}{\tilde q}\right)^{\lambda} + O\left(t^{-2(\lambda-\kappa)/\kappa}\right).
\end{eqnarray}
In summary, the second order of the approximation \eqref{eq:APPIxtSeclastLAST} is given by either or both of
\begin{eqnarray}
t^{-(1+\xi)/\kappa}  C_{2,g} \frac{\partial^{\xi}B_{\kappa}(X_{\kappa})}{\partial(iX_{\kappa})^{\xi}}  \mathrm{e}^{it\varepsilon(p^{*}) - ip^{*}x},\\
t^{-(1+\lambda-\kappa)/\kappa} C_{2,\varepsilon} \frac{\partial^{\lambda}B_{\kappa}(X_{\kappa}) }{\partial(iX_{\kappa})^{\lambda}} \mathrm{e}^{it\varepsilon(p^{*}) - ip^{*}x}.
\end{eqnarray}
Therefore, as long as either $g'(p^{*})\neq0$ or $\varepsilon^{(\kappa+1)}(p^{*})\neq0$ holds, {\it i.e.}, $\xi$=1 or $\lambda=\kappa+1$, the second order of the approximation \eqref{eq:Ifuncapprox} decays as $t^{-2/\kappa}$ in the space-time scaling limit, and hence we obtain Eq.~\eqref{eq:Ifuncapprox}.
\clearpage

\bibliographystyle{iopart-num}
\bibliography{AY_JSTAT_Bib_revised1013_add}

\end{document}